\documentclass[useAMS,usenatbib]{mn2e}

\usepackage{times}
\usepackage{mathptmx}
\usepackage{graphicx}

\usepackage{graphicx}
\usepackage{color}
\usepackage[fleqn]{amsmath}

\title[Rotating triaxial stellar systems]{Models of cuspy triaxial stellar
systems. IV: Rotating systems}

\author[D. D. Carpintero and J. C. Muzzio]
{D. D. Carpintero$^{1,2}$,
\thanks{E-mail: ddc@fcaglp.unlp.edu.ar}
and J. C. Muzzio$^{1,2}$,
\thanks{E-mail: jcmuzzio@fcaglp.unlp.edu.ar}\\
$^{1}$Facultad de Ciencias Astron\'omicas y Geof\'\i sicas,
Universidad Nacional de La Plata, La Plata, Argentina\\
$^{2}$Instituto de Astrof\'\i sica de La Plata (CONICET La Plata--UNLP)}

\begin{document}

\date{}

\pagerange{\pageref{firstpage}--\pageref{lastpage}} \pubyear{}

\maketitle

\label{firstpage}

\begin{abstract}

We built two self--consistent models of triaxial, cuspy, rotating stellar
systems adding rotation to non-rotating models presented in previous papers of
this series. The final angular velocity of the material is not constant and
varies with the distance to the center and with the height over the equator of
the systems, but the figure rotation is very uniform in both cases. Even though
the addition of rotation to the models modifies their original semiaxes ratios,
the final rotating models are considerably flattened and triaxial. An analysis
of the orbital content of the models shows that about two thirds of their orbits
are chaotic yet the models are very stable over intervals of
the order of one Hubble time. The bulk of regular orbits are short axis
tubes, while long axis tubes are replaced by tubes whose axes lie on the
short--long axes plane, but do not coincide with the major axis. Other types of
regular orbits that do not appear in non--rotating systems, like horseshoes and
orbits that cross themselves, are also found in the present models. Finally, our
frequency maps show empty regions where studies of orbits on fixed potentials
found orbits, a likely consequence of the self--consistency of our models that
excludes them.

\end{abstract}

\begin{keywords}
Galaxies: elliptical and lenticular, cD -- Galaxies: kinematics and dynamics -- 
methods: numerical -- Physical data and processes: chaos.
\end{keywords}

\section{Introduction}

It is relatively easy to obtain self-consistent models of
spherical or disk--like stellar systems with simple numerical (e.g., King models)
or even analytical tools (e.g., Schuster or Plummer models), as explained
in textbooks like the one by \cite{BT08}. Models of elliptical galaxies are much
more difficult to build, however, as there is observational evidence, both
statistical \citep[see, e.g.,][]{R96} and on individual galaxies \citep[see,
e.g.,][]{SEPB04}, that shows that at least some ellipticals are triaxial,
and full fledged 3--D models demand resorting to special techniques. Besides,
surface brightness studies of ellipticals tend to show central cusps \citep*[see,
e.g.,][]{Cea93,MSZ95}, that reveal the presence of mass concentrations and probably
black holes, and triaxial and cuspy potentials favor the appearance of chaotic
orbits \citep[see, e.g.,][]{SK00,KS03} further complicating model building for
those objects.

Two main methods are employed to build self--consistent models
of elliptical galaxies: the one due to \cite{S79} and the $N$--body method, originally
proposed by \cite{SS87} to build a bar and later on applied to ellipticals by
\cite*{VKS02}. The former chooses a potential--density pair, builds a library of orbits
in that potential and determines the fraction of each type of orbit needed to obtain
the corresponding density. The $N$--body method adopts an initial distribution of
point masses and integrates the equations of motion until an equilibrium
distribution is reached; a smooth and constant potential is then fitted to
that distribution and the positions and velocities of the bodies are used to
investigate the orbits in that potential.

\cite{S93} himself noted the difficulty to include chaotic orbits
in models built with his method, which turned out to evolve over intervals of the
order of a Hubble time, and the problem became even more serious when cuspy
models were adopted \citep[see, e.g.,][]{MF96}. Although it was suggested that
models containing chaotic orbits could not be stable \citep[e.g.,][]{SK00},
perfectly stable models with large fractions of chaotic orbits were built with
the $N$--body method \citep{VKS02, KV05, MCW05, AMNZ07}, including
cuspy ones \citep*{MNZ09}. In fact, \cite{MCW05} argued that there was no physical
constrain to build self--consistent stable models with chaotic orbits and that
the problem dealt with the method of \cite{S79} itself. A comprehensive recent
discussion of this matter can be found in the paper by \cite{VA12}.

The present series of papers uses the $N$--body method to build
models of cuspy triaxial stellar systems and investigate their stability and
orbital content, both regular and chaotic. In our first paper, \cite{ZM12}
built models resembling E2, E3, E4 and E5 galaxies and they showed that they
were extremely stable over intervals of the order of a Hubble time, even
though they contained fractions of chaotic orbits that exceeded 75 per cent.
The regular orbits of those models were studied, in our second paper, by
\cite*{MNZ13}, and they found that most
of those orbits were short--axis tubes, that the fraction of long--axis
tubes decreased from the E2 through the E5 models and that most of the boxes
were resonant orbits, i.e., boxlets. A curious puzzle was posed by the work
of \cite{HMSH01} who, using the $N$--body method, built cuspy triaxial models
with essentially no chaotic orbits, so that we reexamined their investigation
in the third paper of our series \citep*{CMN14}. We found that their discrepancy
with the other $N$--body works had two causes: their use of a poor method to
detect chaos and a velocity distribution in their model much more isotropic
than that of other authors (low angular momentum orbits are more likely to be
chaotic).

Now, all our previous models are either non--rotating or
have exceedingly slow figure rotation, but rotation is a key ingredient
of the dynamics of elliptical galaxies, and the pioneer works of \cite{BC75} and
\cite{I77} showed that they have angular momentum, even though the
resulting rotation is small enough that in most cases the ellipticity of those
galaxies cannot be attributed  to the rotation itself.

All the investigations performed using the $N$--body method found large
fractions of chaotic orbits in the non--rotating or very slowly rotating
systems, and the comparison done by \cite{M06} of his rotating model with the
same non--rotating one of \cite{MCW05} suggests that rotation should increase
those fractions.  The works of \cite{S82}, and \cite*{DVM11}
are among the few in which a rotating model is used to study the dynamics of
elliptical galaxies, but the former does not mention chaotic orbits and the
latter does not accept that they could contribute significantly to a stable
stellar system. The difficulties of the method of Schwarzschild to accomodate
chaotic orbits have been mentioned above; besides,
\cite{DVM11} do not use a self consistent model, but only investigate orbits in
a rotating triaxial generalization of the potential of \cite{D93}. The recent
work of \cite{VA15} is a welcome addition to the subject but, although they
indicate the presence of chaoticity in their models and that in general it
increased with pattern speed, the information they provide on chaos is rather
scanty.

Thus, we decided to try to build self--consistent, rotating, 
cuspy, triaxial models using the $N$--body method, in order to find out the degree
of chaos that such models harbour and the distribution of the regular orbits, and
that is the subject of the present paper.  The following section describes the
numerical techniques we used to build our models, and their stability is investigated
in Section 3. Section 4 describes the matter and figure rotation of the models, and
Section 5 analyzes its orbital composition, both chaotic and regular. Finally, our
conclusions are presented in Section 6.

\section{Building rotating models}

In order to create stable, cuspy, rotating triaxial models, we take the already
stable, cuspy and triaxial models dubbed E2a and E5a of the first paper of this
series \citep{ZM12}, containing each one $N \simeq 10^6$
bodies. In these models, and in the rest of this work, the gravitational
constant $G=1$, the total mass $M=1$ and the crossing time $T_{\rm cr}$ is
equivalent to about 1/200 of the Hubble time. The slope $\gamma$ of the cusp in
both models, computed as the slope of the $\log\rho(r)$  vs. $\log r$ line for
the innermost 10,000 particles binned in bins of 100 particles each, is $\gamma
\simeq -1$. The semiaxes $a>b>c$ obtained from the 80 per cent most tightly
bound particles are in the ratios $b/a=0.877$, $c/a=0.826$ for the E2a model,
and $b/a=0.814$, $c/a=0.515$ for the E5a model. The triaxiality of the E2a model
is $T\equiv (a^2-b^2)/(a^2-c^2)=0.73$, whereas that of the E5a model is
$T=0.46$. 

We perform the $N$--body integrations with the
self--consistent field (SCF) code of \cite{HO92}, the same used by \cite{ZM12}
to build their models. The code solves the Poisson's equation by expanding the
density and the potential in a set of basis functions chosen in such way that
the lowest order term corresponds to the model of \cite{H90}. The motion of the
bodies is followed with a time--centered leapfrog algorithm that keeps time
reversibility. \cite{ZM12} performed several tests to choose the number of
radial and angular terms in the expansion to finally adopt $n=6$ radial and
$l=4$ angular terms for their models, and those are the same numbers we use
here.

To quantify the angular momentum
acquired by each model, we computed the specific angular momentum
introduced by \cite{P71}:
\begin{equation}
\lambda = \frac{L\sqrt{|E|}}{G M^{2.5}},
\end{equation}
where $L$ and $E$ are the total angular momentum and energy of the
model, respectively.

In order to add rotation while maintaining the triaxiality and
the 'cuspiness' of the models, we proceeded in the following
way. First, we gave to each model a certain amount of rotation
by adding to each particle at position $(x_i,y_i,z_i)$ and with velocity $(\dot
x_i,\dot y_i, \dot z_i)$ an angular velocity $\Omega\, {\bf e}_z$, i.e.,
\begin{align}
\dot x'_i&=\dot x_i-\Omega\, y_i, \nonumber \\
\dot y'_i&=\dot y_i+\Omega\, x_i, \\
\dot z'_i&=\dot z_i. \nonumber 
\end{align}

Now, this simple recipe has a major flaw: it changes the
energy of the model, which should be kept in order to
maintain as much as possible its other characteristics. One
way to procceed is to substract from the velocity modulus of each particle an
amount necessary to conserve the total energy after applying
the rotation. The substracted fraction $0<k<1$ should be the
same for each  particle in order to keep the kinematical structure of the
model. One then has, equalizing the initial and final kinetic energies,
\begin{multline}
\frac{1}{2}\sum_{i=1}^N m_i ( \dot x_i^2+\dot y_i^2)=\\ 
\frac{1}{2}\sum_{i=1}^N m_i\left[ k^2(\dot x_i^2+\dot y_i^2) + 
2k\Omega(x_i \dot y_i-y_i \dot x_i)+
\Omega^2(x_i^2+y_i^2)\right],
\end{multline}
where $m_i$ is the mass of the $i$-th particle, and $N$ is the
number of particles. Solving for $k$, it may happen that the resulting quadratic
polynomial has imaginary roots. This means that the amount of rotational energy
injected into the system is so large that it doesn't allow to recover the
initial kinetic energy (that is, there are particles which
must have less than zero velocity to achieve that). In any
other case, since the linear coefficient $\sum_i m_i\Omega(x_i \dot y_i-y_i \dot
x_i)$ is approximately zero for a non-rotating system, we expect that the two
roots be opposite each other. Moreover, calling $a$ the
coefficient that multiplies $k^2$ and $c$ the independent term, it can be
easily seen that $|c|<|a|$, so that the moduli of the roots
will be less than 1. Thus, except for the imaginary case, a unique value $0<k<1$
is expected among the roots, as was always the case. With the
value of $k$, the new velocities are computed by means of
\begin{align} 
\dot x'_i&=k\dot x_i-\Omega\, y_i, \nonumber \\
\dot y'_i&=k\dot y_i+\Omega\, x_i, \\
\dot z'_i&=k\dot z_i.\nonumber 
\end{align}

As already said, with this procedure only a maximum amount of
angular velocity can be applied to a given model. For example, the rotation that
could be added to the E2a model amounted to $\Omega=0.59495$, whereas that of
the E5a model was $\Omega=0.85870$. Thus we tried to add more
angular momentum in the following way.

We first let each model relax for 50 $T_{\rm cr}$, using the
SCF code. After that, we aligned the coordinate axes with the semiaxes of
the inertia tensor of the 80 per cent most tightly bound
particles of each model, so that the $x$ axis coincided with the semimajor axis
and the $z$ axis with the semiminor axis, and eliminated the particles with
positive energy and the 2 per cent of the remainder that have energies closer to
zero.

Then, we added a new amount of angular velocity. Since to do
this we cannot follow the abovementioned procedure (a
fraction of the velocities of the particles are already part of the angular
motion, which we do not want to substract), we included in the
computation the old angular velocity, dubbed now $\Omega_0$, yielding
\begin{multline}
\frac{1}{2}\sum_{i=1}^N m_i ( \dot x_i^2+\dot y_i^2)= \\
\frac{1}{2}\sum_{i=1}^N m_i\left\{ k^2(\dot x_i^2+\dot y_i^2) + 
2k\Omega(x_i \dot y_i-y_i \dot x_i)- \right.\\
\left.\left[(\Omega_0^2-\Omega^2)(x_i^2+y_i^2)+
2\Omega_0(x_i \dot y_i-y_i \dot x_i)
\right]\right\}.
\label{cambiodos}
\end{multline}

Now the quadratic polynomial can have roots of different moduli and, in
principle, with the same sign. Nevertheless, physically we expect that the
reduction in energy by decreasing the velocities should be possible (so two
negative roots are not expected), and that only one way to do it is feasible (so
two positive roots are not expected either). Thus, the resulting roots should be
one positive and one negative (which was always the case). Once the velocities
have been reduced again using Eq. (\ref{cambiodos}), the whole procedure is
repeated (including the rotation of the axes and the relaxation) until the
amount of angular 
motion to be 
added cannot be made much different from zero.

For the E2a model, the second introduction of angular momentum was the last to
produce a significant growth of $\lambda$, whereas in the case of the E5a
model, it was the third one. The final values were $\lambda=0.124$ for the E2a
model, and $\lambda=0.178$ for the E5a model.

Then, we let the models relax during 200 $T_{\rm cr}$ (about a Hubble time); the
final snapshot was again rotated using the moments of the inertia tensor of the
80 per cent most tightly bound particles. These last systems
(one for each model) were
evolved during an additional interval of 300 $T_{\rm cr}$ in order to verify
their stability. For each of the two
models, the snapshot corresponding to 100 $T_{\rm cr}$ of this last evolution
was the one used to study and classify the respective orbital contents; we call
them E2af and E5af, respectively, for brevity.

\begin{figure*}
\resizebox{\hsize}{!}{\includegraphics{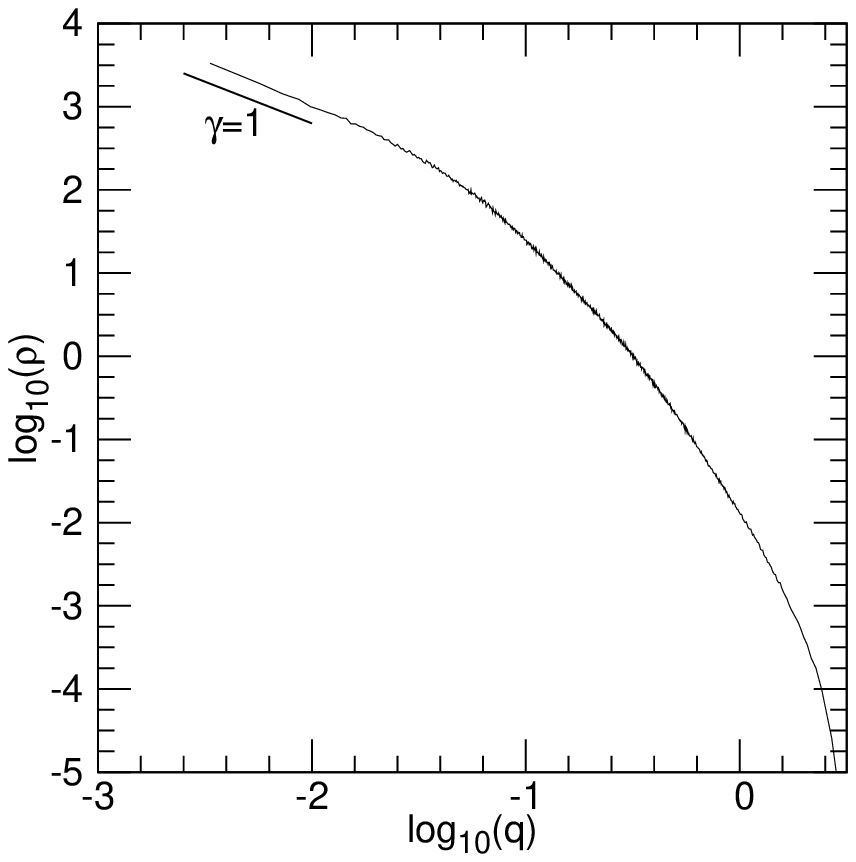}\hspace{1cm}
                      \includegraphics{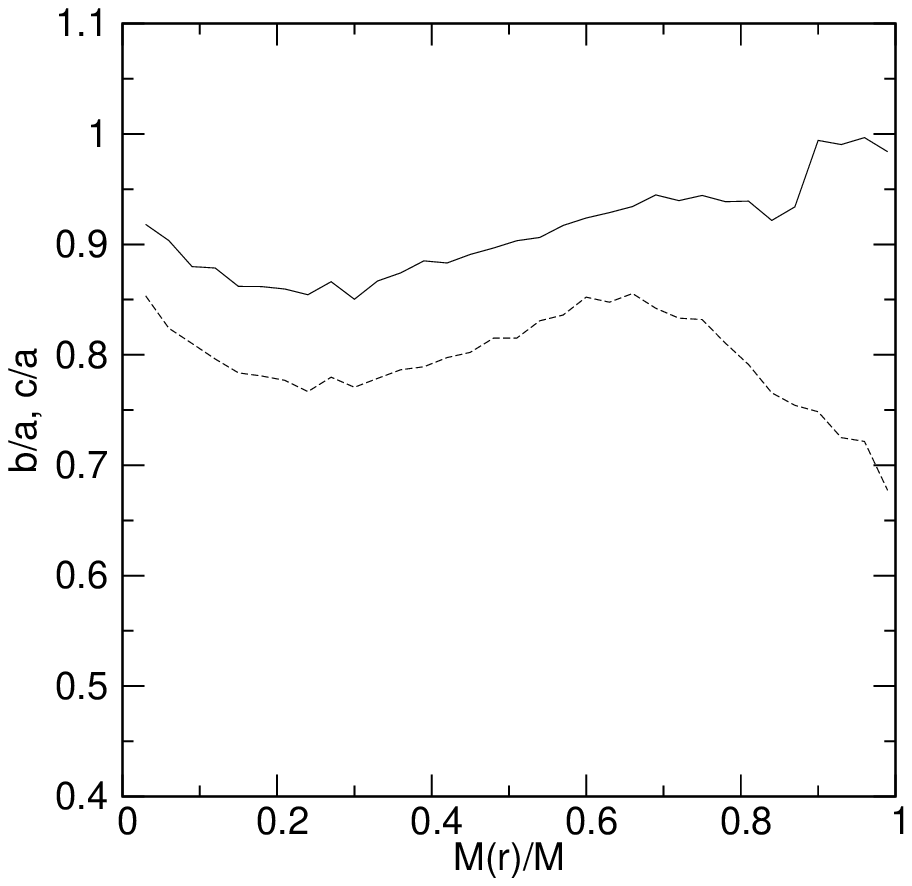}}
\resizebox{\hsize}{!}{\includegraphics{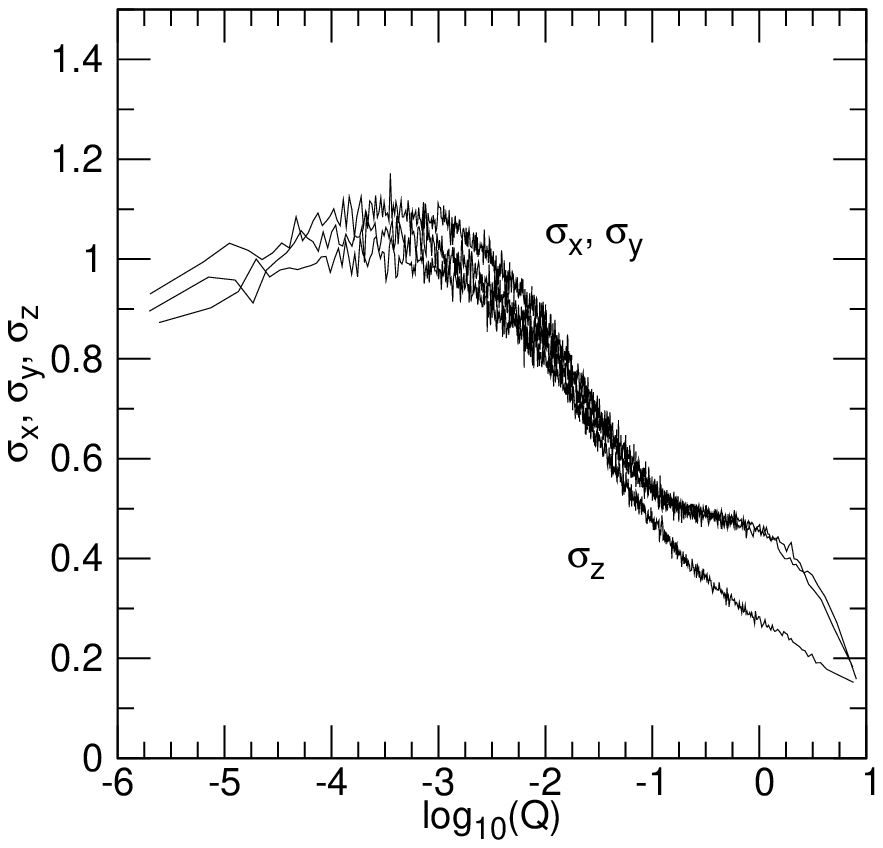}\hspace{1cm}
                      \includegraphics{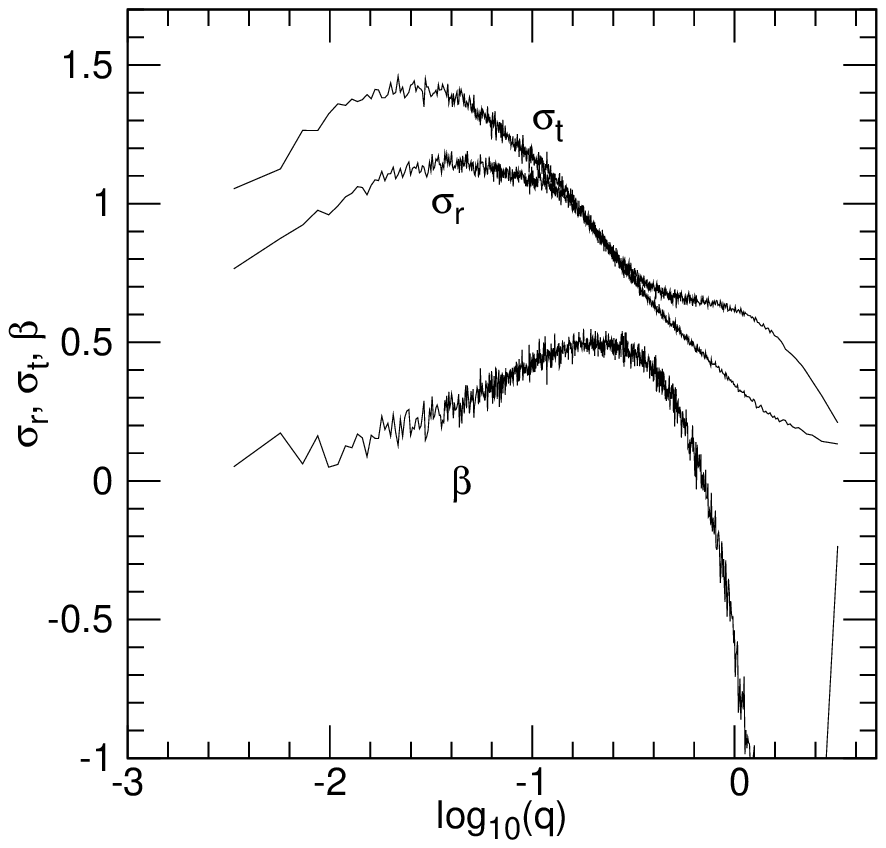}}
\caption{Structural and kinematical properties of model E2af; $q$ is the
ellipsoidal radius, $Q$ the 
projected ellipsoidal radius, and $M(r)/M$ is the mass interior to $r$
in units of the total mass. Upper left: density profile; a short
segment with slope $-1$ was added for reference. Upper right: axial ratios; the
upper curve corresponds 
to $b/a$ and the lower one to $c/a$. Lower left: projected velocity dispersions
along the main axes;
upper, intermediate and lower curves correspond, respectively, to $\sigma_{x},
\sigma_{y}$ and $\sigma_{z}$. 
Lower right: tangential and radial velocity dispersions and velocity anisotropy
parameter $\beta$.} 
\label{ge2a}
\end{figure*}

\begin{figure*}
\resizebox{\hsize}{!}{\includegraphics{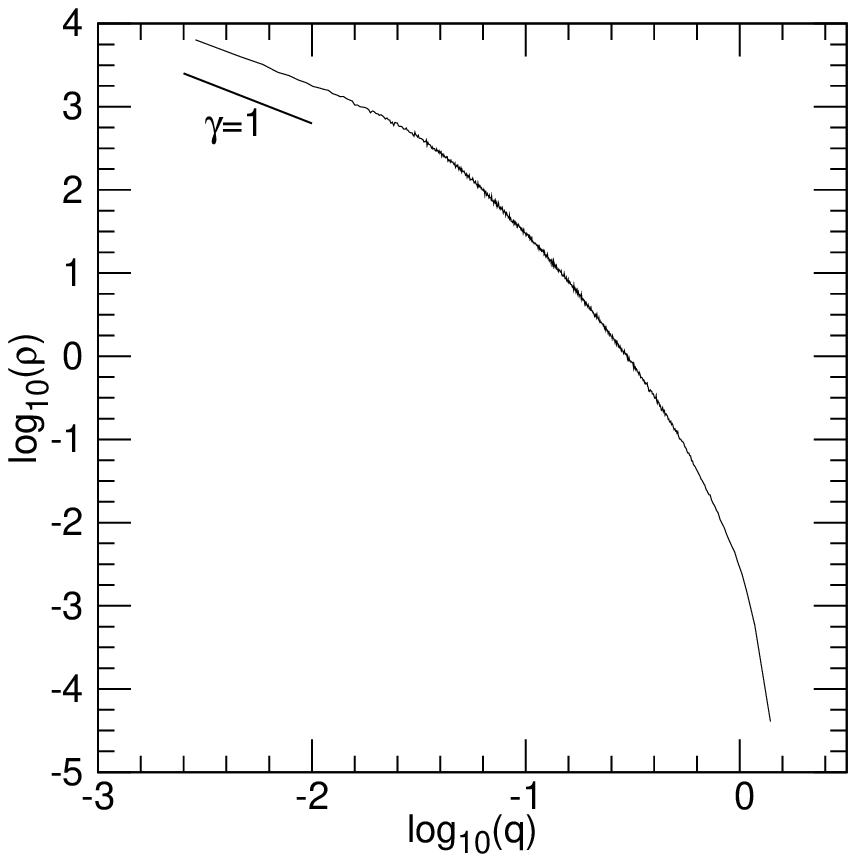}\hspace{1cm}
                      \includegraphics{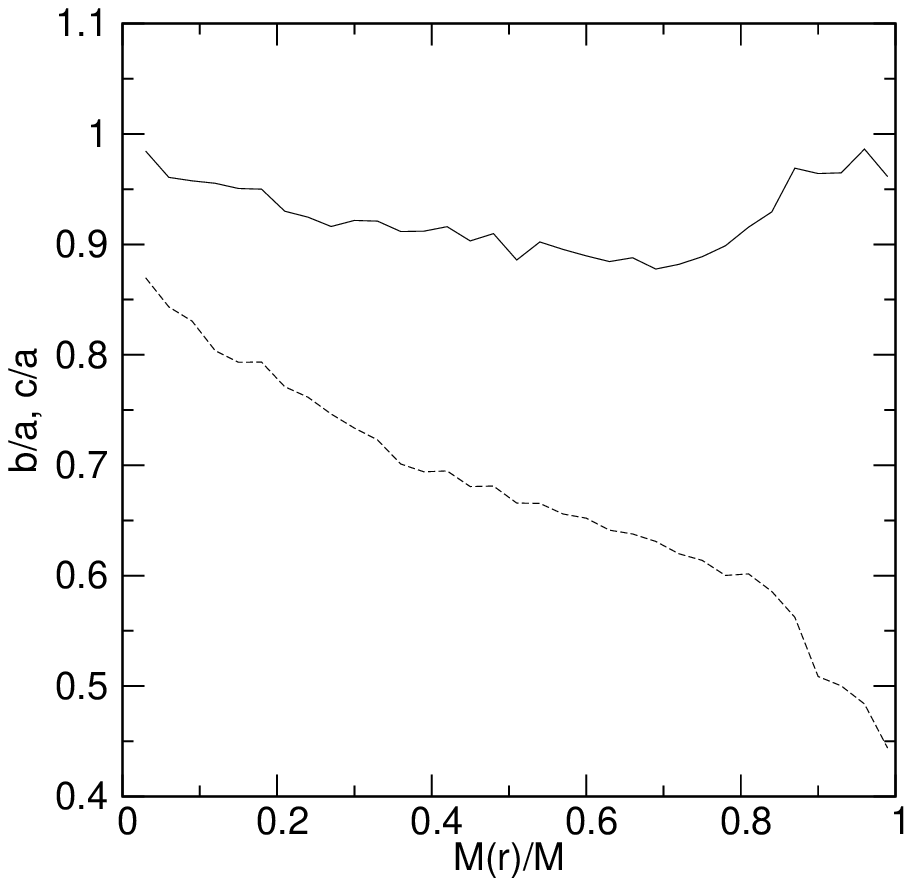}}
\resizebox{\hsize}{!}{\includegraphics{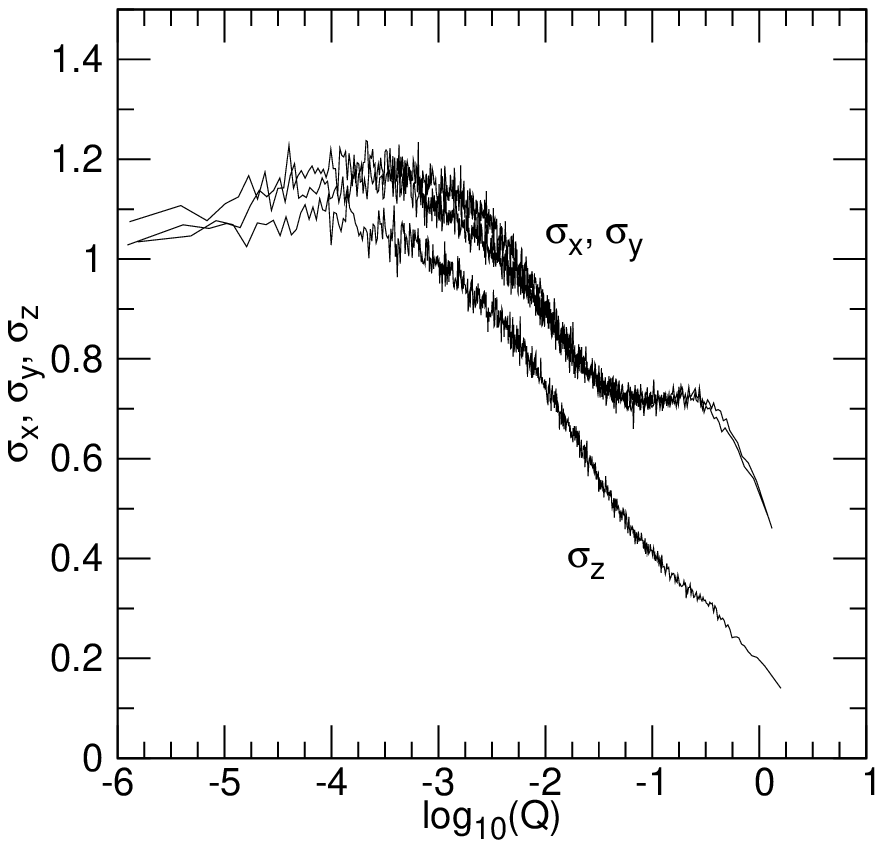}\hspace{1cm}
                      \includegraphics{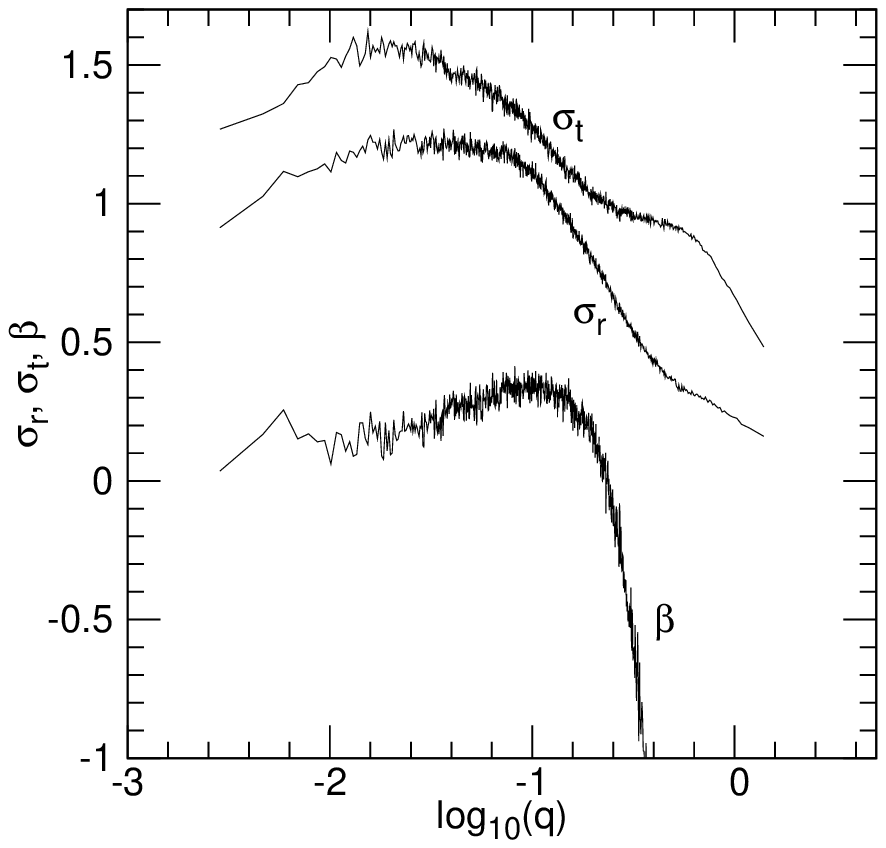}}
\caption{Same as Fig. \ref{ge2a} but for model E5af.} 
\label{ge5a}
\end{figure*}

Fig. \ref{ge2a} shows the kinematic and structural characteristics of the
E2af model. In this Figure, the abscisae $q$ stands for the
ellipsoidal radius 
\begin{equation}
q=\left(\frac{x^2}{a^2} + \frac{y^2}{b^2} + \frac{z^2}{c^2}\right)^{1/2},
\end{equation}
where $a$, $b$ and $c$ are the semiaxes of the model, computed following
the method of \cite{DC91}, as adapted by \cite{HMSH01}; $Q$, on the
other hand, is the
same as $q$ but projected onto the coordinate planes. The abscisae $M(r)/M$ is
the mass interior to radius $r$
in units of the total mass of the model. The upper left panel shows
the density profile of the model, computed using ellipsoidal shells with
constant axial
ratios taken at the half-mass radius. As can be seen, the rotation has not
substantially affected the cusp of the model ($\gamma = -1.023 \pm 0.020$,
computed from the densities of the 10,000 innermost particles binned in 100
intervals of distance to the center). The upper right panel shows the
axial ratios, computed in shells each containing 3 per cent of the particles,
previously sorted by energy. The ratios of the semiaxes of the
shell containing 81 per cent of the mass in its interior are $b/a=0.939$ and
$c/a=0.791$, in good agreement with the values that correspond to the 80 per
cent most tightly bound bodies ($b/a=0.925$, $c/a=0.823$, $T=0.447$), i.e.,
the parameters used in our previous works. Comparing these values with those of the original
E2a model, we can see that the $b/a$ ratio has increased, whereas
the $c/a$ ratio and the triaxiality have decreased.
The bottom panels show the
kinematical profiles, in which $\sigma_i^2=\langle v_i^2 \rangle - \langle v_i
\rangle^2$, where $i$ is a Cartesian direction $x$, $y$ or $z$ in the left panel,
and $r$ (radial) or $t$ (tangential) direction in the right panel. The
anisotropy 
parameter $\beta=1-\sigma_{\rm t}^2/(2\sigma_{\rm r}^2)$. Fig. \ref{ge5a} shows the
same characteristics but for the E5af model. Again, the cusp is well preserved
($\gamma=-0.956 \pm 0.023$). In this case, for the same shell as
before, $b/a=0.916$, $c/a=0.602$, while for the 80 per cent most tightly bound
particles, $b/a=0.896$, $c/a=0.641$ and $T=0.335$, i.e., both axial ratios have
increased and the triaxiality decreased with respect to the E5a model.
It might seem odd that rotation has not made the model flatter,
but it should be recalled that the process that led from the non--rotating to the
rotating model implied considerable changes in structure of the former. For
example, about 30 per cent of the original particles were eliminated because
they increased their energies beyond the established limits, and the $a$ semiaxis
was reduced to about half its initial size.

In order to estimate the meaning of our length unit (l.u.) and
time unit (t.u.) for real galaxies, we computed the effective radii $R_{\rm e}$
from the $(x, z)$ projection (0.142 and 0.0794 l.u., respectively, for the E2af
and E5af models) and the central radial velocity dispersion $\sigma_0$ from the
$y$ components of the velocities of the $10,000$ particles closer to the centre
on that projection (0.987 and 1.099 l.u./t.u., respectively, for the E2af and
E5af models). As in our previous work \citep{AMNZ07, MNZ09, ZM12}, we chose for
comparison galaxies NGC1379 and NGC4697 \citep{NCR05, FP99},  whose
mass-to-light ratio gradients are zero. Comparing their observed values of
$R_{\rm e}$ (2.5 and 5.7 kpc, respectively) and $\sigma_0$ (128 and 180 km
s$^{-1}$, respectively) with those from our models,  we conclude that values
between about 18 and 72 kpc can be used as our length unit and  values between
about 0.14 and 0.44 Gyr as our time unit. Then, the Hubble  time can be
estimated as between 32 and 100 t.u., or between 180 and 290 $T_{\rm cr}$, and
we will adopt a value of 250 $T_{\rm cr}$, hereafter.

\section{Stability of the models}

We used the final 300$T_{\rm cr}$ runs to check for systematic changes in the
density $\rho_{\rm c}$ of the innermost 10,000 particles, the moments of inertia
$X$, $Y$ and $Z$ for each axis, and $\gamma$. To improve the precision, straight
lines were fitted to the results obtained every 25$T_{\rm cr}$ and we obtained
the percentage variation corresponding to 250$T_{\rm cr}$, i.e., one Hubble time
according to our estimate. The results are presented in Table \ref{stable} and
the changes are very small, indeed. Moreover, as we have shown before
\citep{AMNZ07, MNZ09, ZM12},   even those tiny variations are most likely not
real but due to relaxation effects of the $N$--body code \citep{HB90}.

Nevertheless, since our models are rotating, what is conserved is not
the energy of a particle in the fixed reference frame of the $N$--body code,
but its energy in the frame that rotates with the ellipsoidal figure,
i.e., the Jacobi integral (leaving aside the numerical relaxation
effects, of course). Therefore, even though we eliminated the particles
with positive, and slightly negative, energy in the former reference
frame, there are still particles that can escape from the system. As
explained by \cite{BT08}, particles that have values of the Jacobi
integral larger than the value of the effective potential at the
$L_{1}$ and $L_{2}$ Lagrangian points or that are initially outside
the contour through those points, can \emph{in principle} escape from the
system but, due to the protective effect of the Coriolis force, that
does not mean that those particles will \emph{necessarily} escape. We
obtained the limiting values of the Jacobi integral as $-1.113$ and
$-2.042$, respectively, for the E2af and E5af models, and found that
123,417 (or 15.6 per cent) and 140,252 (or 20.5
per cent) particles, respectively,
exceeded those limits. Nevertheless, after the 300$T_{\rm cr}$ evolution,
only 7 (or 0.00088 per cent) and 112 (or 0.016
per cent) particles from the E2af 
and E5af models, respectively, were found at a distance from the center
that exceeded that of the initially most distant particle in the corresponding
model. Thus, we can conclude that our models are extremely stable over
intervals of the order of one Hubble time.

\begin{table*}
\centering
\caption{Percentage variations over one Hubble time.} 
\label{stable}
\begin{tabular}{@{}lccccc@{}}
\hline
Model & $\rho_{\rm c}$ (\%)& $X$ (\%)& $Y$ (\%) & $Z$ (\%) & $\gamma$ (\%)\\ 
\hline 
E2af  & $-0.78\pm 0.89$ & $-0.86\pm 0.21$ & $0.91\pm 0.28$ & $0.15\pm 0.29$ &
 $\hphantom{-}1.47\pm 1.20$\\ 
E5af  & $-0.12\pm 1.04$ & $-1.62\pm 0.18$ & $1.52\pm 0.23$ & $0.54\pm 0.23$ &
 $-1.60\pm 1.43$\\
\hline
\end{tabular} 
\end{table*}

\section{Figure rotation versus matter rotation}

\begin{figure*}
\resizebox{\hsize}{!}{\includegraphics{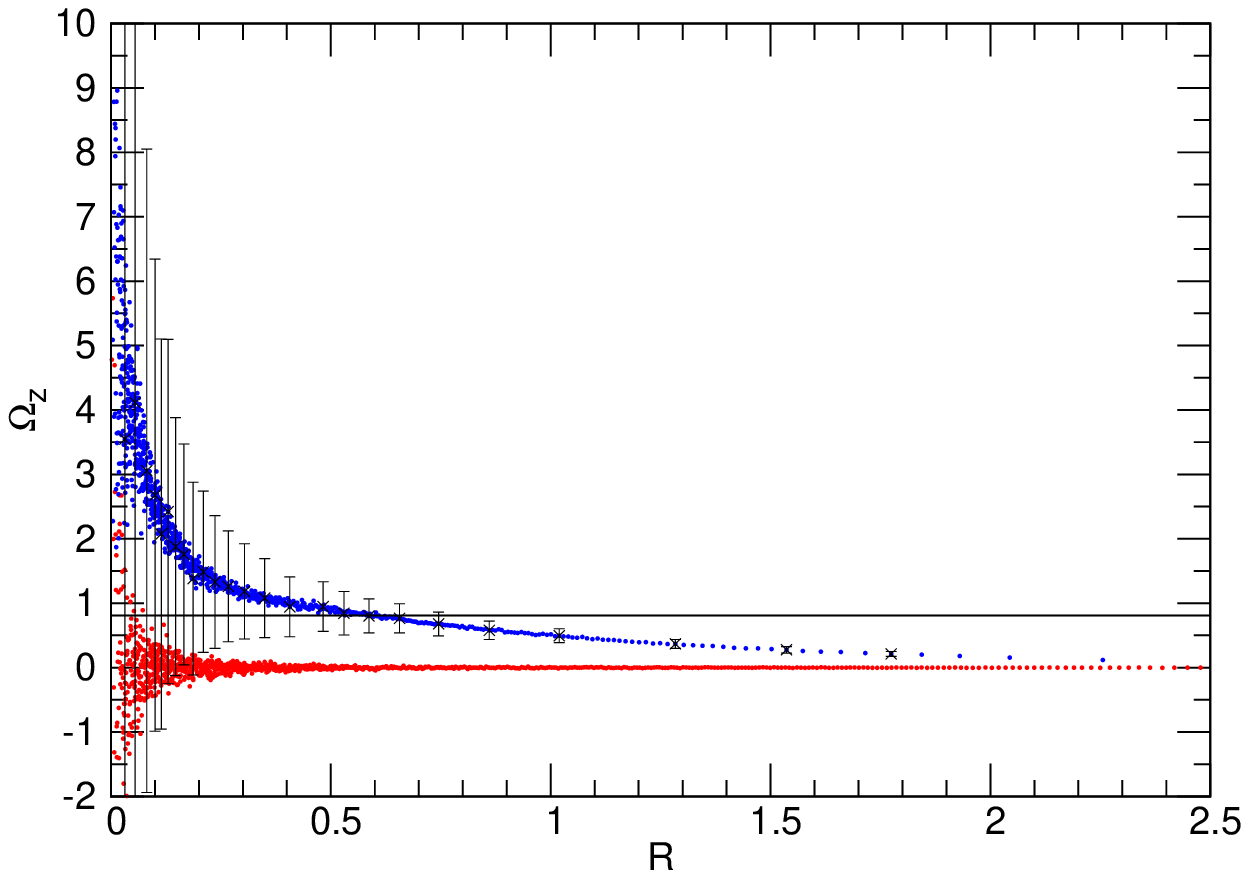}\hspace{1cm}
                      \includegraphics{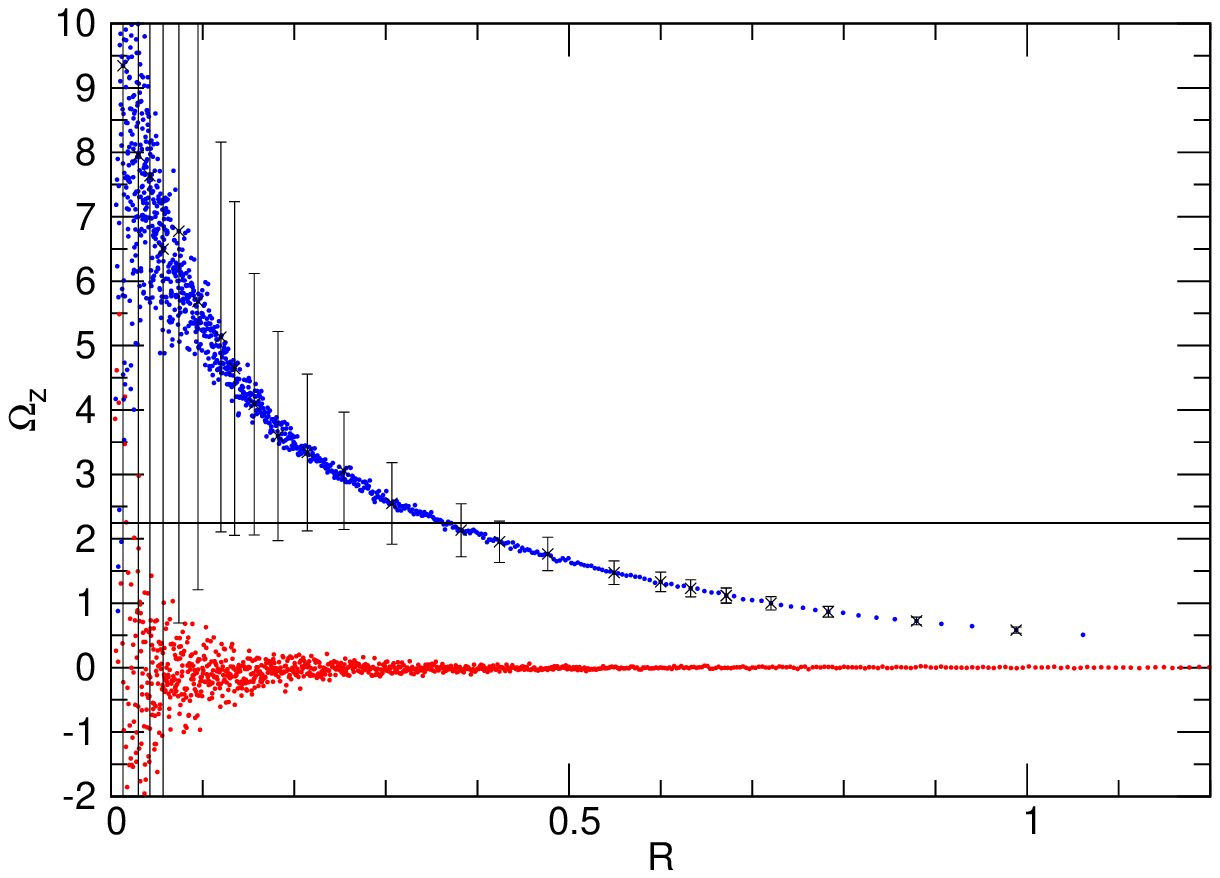}}
\caption{Left: Angular velocity $\Omega_z$ as function of the cylindrical radius
$R=\sqrt{x^2+y^2}$ for the initial E2a model (lower set of points, red in the electronic
version) and for the final E2af model (upper set of points, blue in the electronic version).
Each point corresponds to the mean angular velocity of a cilyndrical shell containing
1/1000 of the particles. The dispersions within some shells are also shown with bars.
A horizontal line at $\Omega=0.8108$ shows the angular velocity of the figure rotation.
Right: the same, but for the E5a and E5af models. The angular velocity of the figure rotation is at
2.2454, marked with an horizontal line. In both cases, the initial models reach $R \simeq 4$.}  
\label{omegas}
\end{figure*}

In a non-rotating stellar system, one expects that the outermost
particles have approximately zero angular velocity. This is because the external
region (halo) of a self-gravitating system in equilibrium, particularly when it
is the result of a gravitational collapse, is expected to contain almost all
radial or quasi radial orbits \citep[e.g.,][and references therein.]{BT08}.
On the other hand, the internal region (core) tends to have particles with a more
isotropic distribution of velocities, meaning that there is a wide dispersion of
angular velocities in every direction, summing up to zero angular velocity for
the system as a whole. Fig. \ref{omegas} shows that that is indeed the
initial state of our E2a and E5a systems (lower sets of points, red in the
electronic version). 

\begin{figure*}
\resizebox{\hsize}{!}{\includegraphics{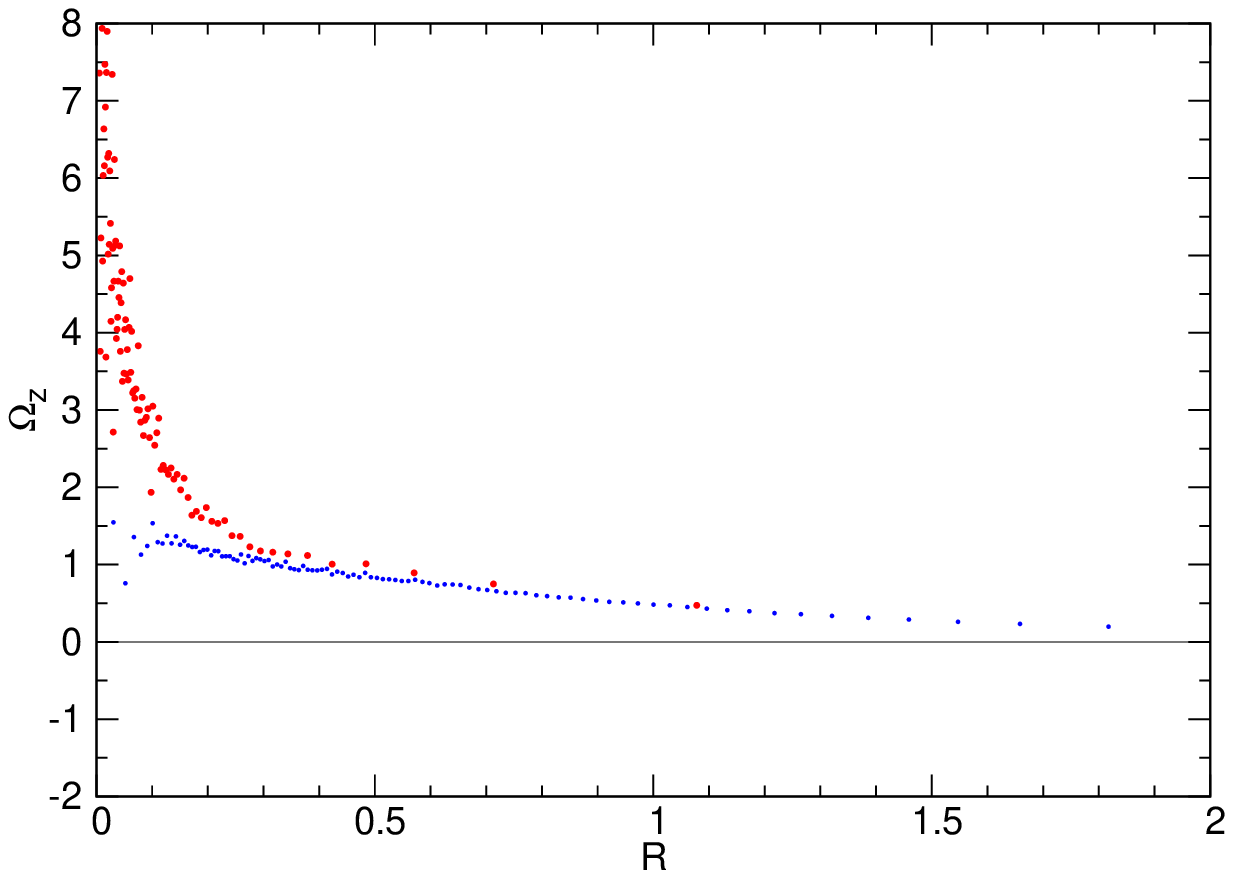}\hspace{1cm}
                      \includegraphics{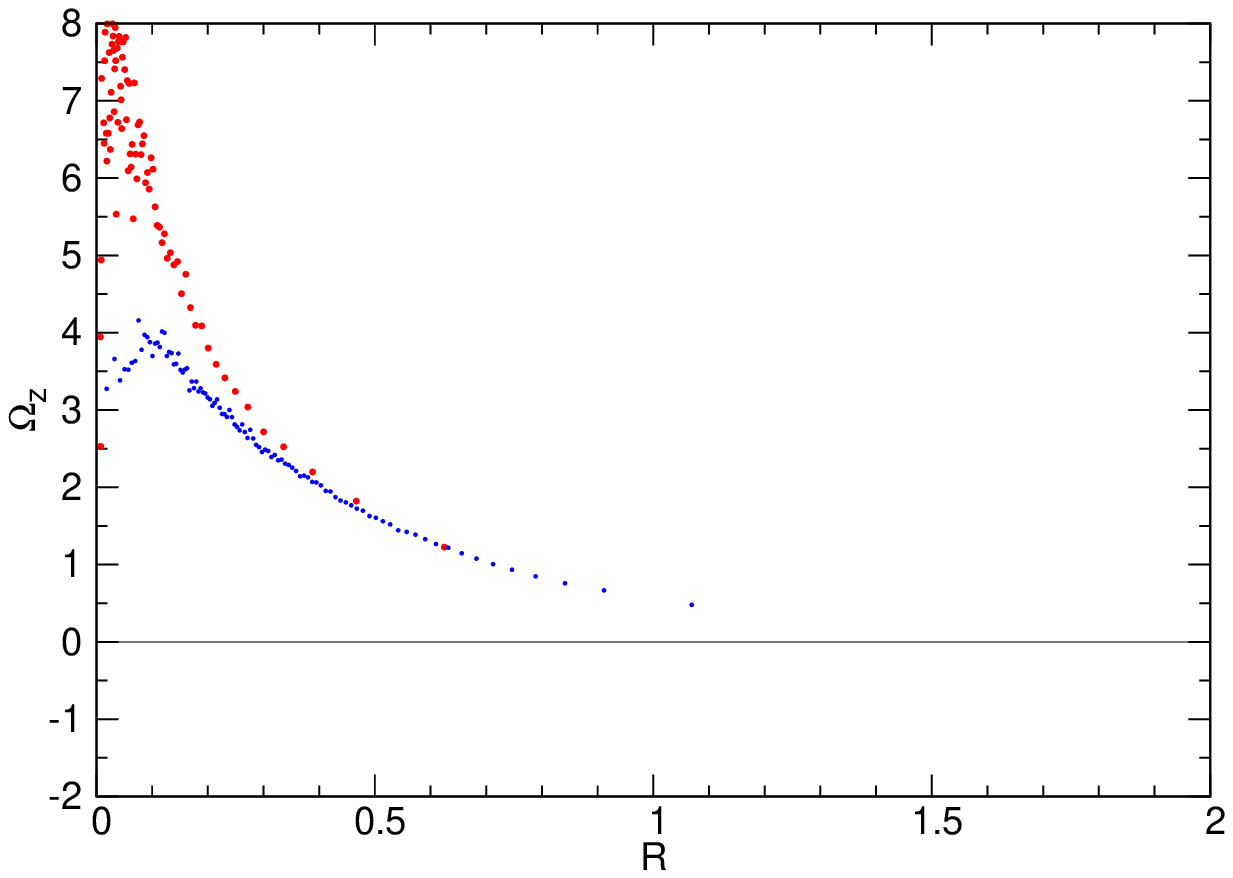}}
\caption{Left: Angular velocity $\Omega_z$ as function of the cylindrical radius
$R=\sqrt{x^2+y^2}$ for
the E2af model, taking only the 20 per cent of particles closest to the
equatorial plane (upper set of points, red in the electronic version) and the 20
per cent farthest from that plane (lower set, blue in the electronic version).
Each point corresponds to the mean angular velocity of a
cilyndrical shell containing 1/100 of the corresponding 20 per cent particles.
Right: the same, but for the E5af model.}  
\label{omegaz}
\end{figure*}

With the first injection of angular momentum, each particle acquires the
same amount of angular velocity, and the resulting profile of $\Omega_z(R)$
(not shown in Fig.\ref{omegas}) is the same as before, but raised by that amount.
Therefore, the energy of most of the halo particles exceeds their binding energy
and they escape from the system when it is let to evolve. Core particles, on the
other hand, tend to remain bound but, as their energy has increased,
many will go to populate the halo, thus decreasing their angular velocity due
to an approximate conservation of angular momentum. Injection of more angular
momentum after this evolution will initially raise again the entire profile without
modifying its shape, and the subsequent evolution will yield a new depletion
of the halo because of escapes, migration of core particles to the halo, and the
lowering of the external profile of $\Omega_z(R)$. The final angular velocity
profile of our models, displaying a strong diferential rotation, is shown in
Fig.\ref{omegas} (upper sets of points, blue in the electronic version).
Moreover, the final rotation also depends on $z$. Fig. \ref{omegaz} shows the
mean angular velocity of cylindrical shells for each final model taking only
the 20 per cent of the particles closest to the $(x,y)$ plane (upper
sets of
points, red in the electronic version) and the 20 per cent farthest from that
plane (lower sets, blue in the electronic version). We recall
that, in physical units, the unit of abcissae in Figs. \ref{omegas} and \ref{omegaz}
corresponds to about 20 to 40 kpc, for model E2af, and to about 30 to 70 kpc,
for model E5af, and the unit of ordinates to about 4 to 7 km
s$^{-1}$ kpc$^{-1}$, for model E2a, and to about 2 to 4 km
s$^{-1}$ kpc$^{-1}$, for model E5af.

\begin{figure*}
\resizebox{\hsize}{!}{\includegraphics{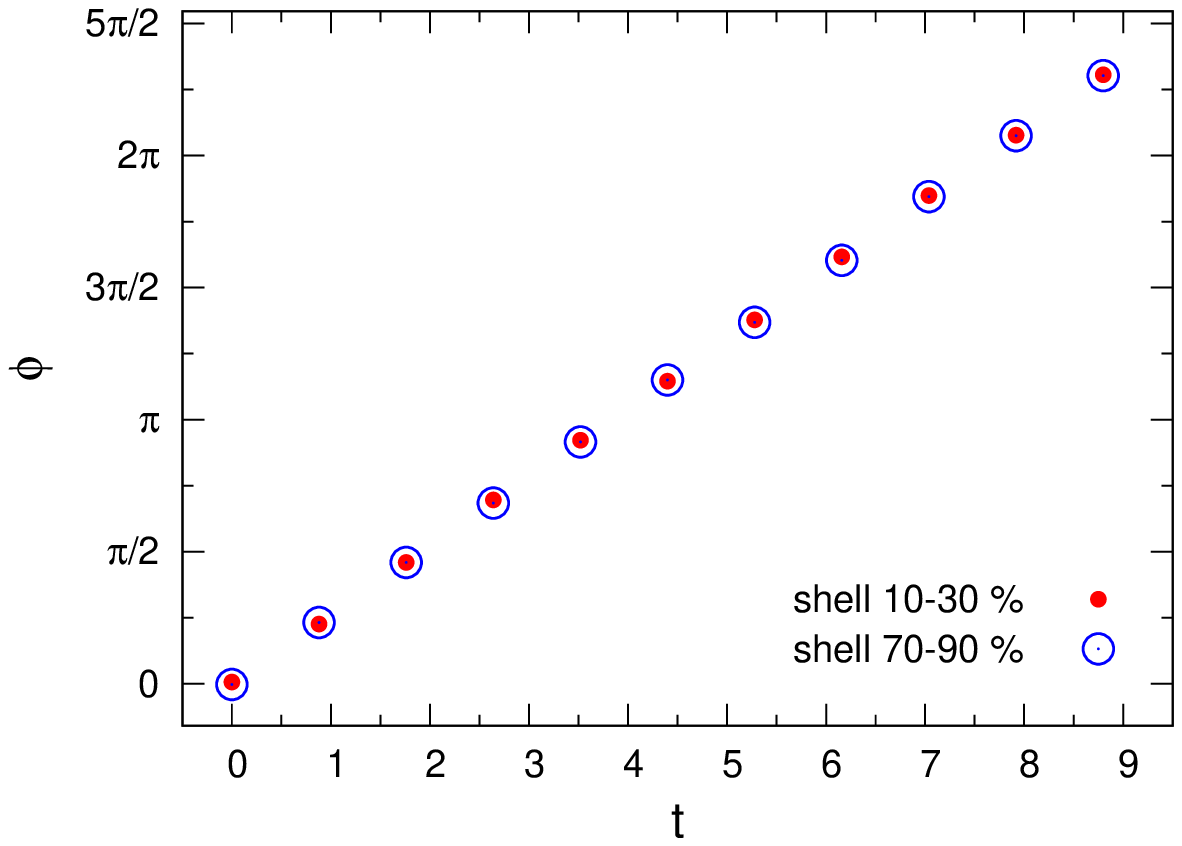}\hspace{1cm}
                      \includegraphics{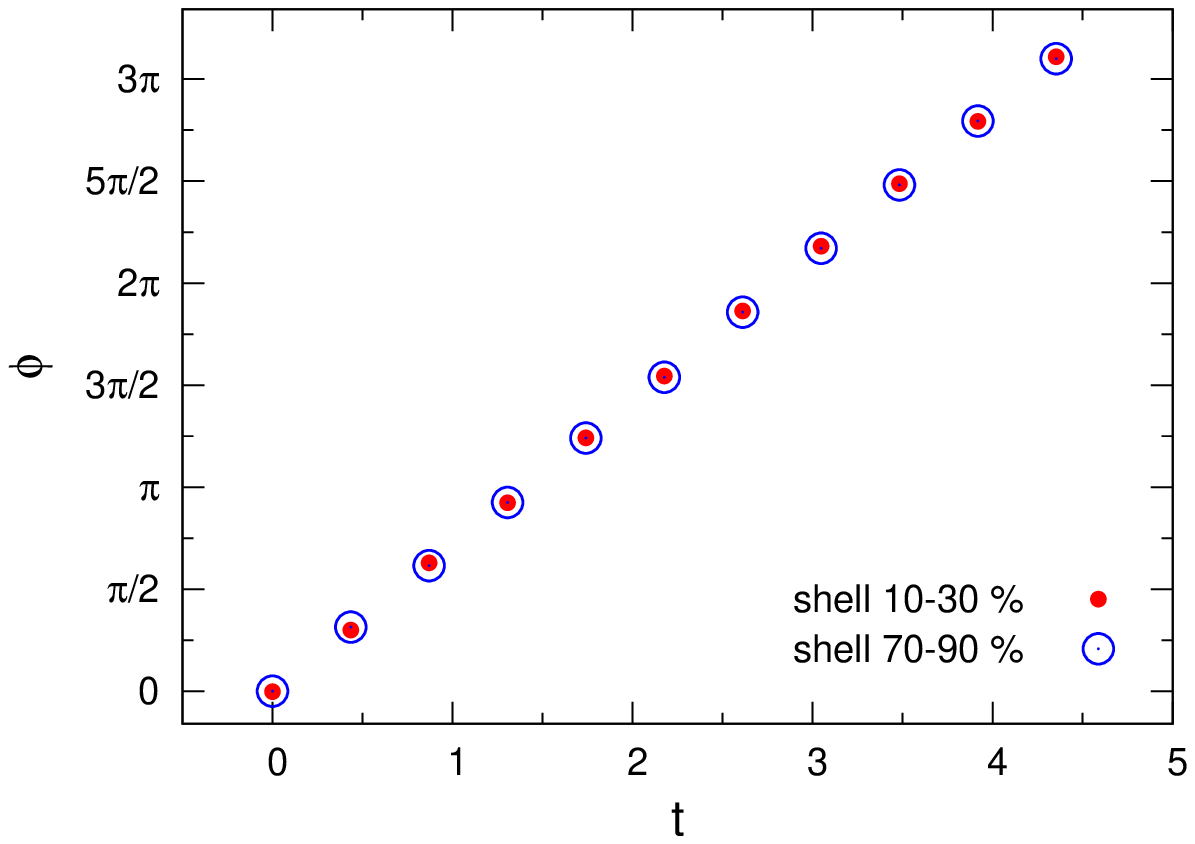}}
\caption{Figure rotation of the E2af (left) and E5af (right) models.
Only the results corresponding to two shells containing
20 per cent of the particles are shown for clarity, but the other shells give
essentially the same results
(see text). The time spans, expressed in t.u., differ for each model due to
the different values of their crossing times, but in both cases they correspond
to $25 T_{\rm cr}$.}
\label{figrot}
\end{figure*}

These results clearly show that, although the amount of angular velocity
imparted to all the particles was the same, the resulting rotation is far from
that of a rigid body: the dynamical evolution generates a complex rotation
pattern in the models. This turns out to be in sharp contrast with the so-called
figure rotation, that is, the rotation of the triaxial shape of the system
around its short axis. To measure this rotation, we let our final models evolve
for an additional interval of 25 $T_{\rm cr}$, obtaining ten snapshots, and we
analyzed each one of these snapshots as follows. First, we sorted all the
bound particles according to their energy and distributed them
in 20 percent bins, from the 10 percent most tightly bound to
the 90 percent most tightly bound, and then we computed for
each bin the moments of the inertia tensor and obtained the angle between the
semimajor axis of the ellipsoid and the positive $x$ axis; the 10 per cent less
tightly bound and most tightly bound
particles were excluded because the corresponding ellipsoids have $b/a \simeq 1.0$
(see Figs. \ref{ge2a} and \ref{ge5a}, upper right frames) and the angle
determination is uncertain. Plots of this angle $\phi$ versus the elapsed time
show the rotation of the ellipsoidal figures, and it was clear
that the different energy bins of the figure of each system are rotating with the
same angular velocity. We present such plots in Fig. \ref{figrot}, but we only
included the innermost and outermost shells for clarity. The angular velocity of
this figure rotation, or pattern velocity, was computed using the 80 per cent most
tightly bound particles and turned out to be $\Omega_{\rm p}=0.8108 \pm
0.0054$ and $2.2454 \pm 0.0094$, respectively, for models E2af and E5af. These
values are shown in Fig. \ref{omegas} with horizontal lines for comparison with
the matter rotation and, in physical units, they imply rotation
periods between 1.0 and 1.7 Gyr for E2af, and between 0.8 and 1.2 Gyr for E5af.

\section{Orbital content of the models}

We investigated the orbital content of our models with the same methods from the
previous papers in this series \citep{ZM12,MNZ13,CMN14}, that is, Lyapunov
exponents to detect chaoticity, and frequency maps to study regular orbits. We
used the numerical tools we had developed for those investigations, but we would
wish to call the atention of readers new to the field to the software tool {\sc
smile} \citep{V13} that they might find useful.

\subsection{Chaoticity}
\label{subchao}

We randomly selected $3,961$ and $3,429$ particles, respectively, from models E2af
and E5af and we took their positions and velocities as the initial values to obtain
the orbits and investigate their chaoticity. The potentials were fixed, keeping
constant the coefficients of their expansions, and the integrations were carried out
in coordinate systems rotating with the corresponding angular velocities. 

Since the potentials were fixed, all our orbits obey the Jacobi integral, and
regular orbits have to obey at least two additional isolating integrals,
but we can have two kinds of chaotic orbits, partially chaotic orbits that obey
only one additional integral, and fully chaotic orbits that have no isolating
integrals other than the Jacobi integral. Regular, partially and fully chaotic
orbits can be classified obtaining the six Lyapunov exponents. Since phase space
volume is conserved, the exponents come in three pairs  of the same absolute
value and opposite sign. The Jacobi integral guarantees that one of those  pairs
is always zero, and each additional isolating integral makes zero  another pair,
so that regular orbits have all their Lyapunov exponents equal to zero,
partially chaotic orbits have one non-zero pair and fully chaotic orbits have
two.

The numerical equivalent of the Lyapunov exponents are the finite time Lyapunov
characteristic numbers (hereafter FT-LCNs). As in our previous works, we
computed them using the LIAMAG subroutine \citep{UP88}, kindly provided by D.
Pfenniger, adopting integration and normalization intervals of 10,000 t.u. and 1
t.u., respectively. We will refer to the largest FT-LCN of a given orbit as
$L_{\rm max}$ and to the second largest one as $L_{\rm int}$, hereafter.  Since
the FT-LCNs are obtained from numerical integrations over a finite time
interval, rather than the infinite one required to obtain Lyapunov exponents,
they cannot reach zero value, but only a limiting minimum value, $L_{\rm lim}$. 
As in our previous work, we used plots of the $L_{\rm int}$ versus $L_{\rm max}$
distribution to estimate a value of $L_{\rm lim}=0.0018\ ({\rm t.u.})^{-1}$, the
same one obtained by \cite{ZM12}. This $L_{\rm lim}$ corresponds to a Lyapunov
time of 556 t.u., which is equivalent to about 5 or 6 Hubble times for our
models, and one might wonder whether it is reasonable to use such a low $L_{\rm
lim}$ to separate regular from chaotic orbits. We have dealt with this matter in
our previous work \citep{AMNZ07, MNZ09,ZM12} and we repeat here the same
analysis done before, using also a limiting value of $L_{\rm lim}=0.0100\ ({\rm
t.u.})^{-1}$ for comparison. First, we separated the orbits of each model into
three groups: a) Those with $L_{\rm max} < 0.0018\ ({\rm t.u.})^{-1}$, i.e.,
those that are classified as regular for both choices of $L_{\rm lim}$ (REGREG,
hereafter); b) Those with  $0.0018\ ({\rm t.u.})^{-1}\leq L_{\rm max} < 0.0100\ 
({\rm t.u.})^{-1}$, i.e., those that are classified as regular  for $L_{\rm
lim}=0.0100\ ({\rm t.u.})^{-1}$, but as chaotic for $L_{\rm lim}=0.0018\ ({\rm
t.u.})^{-1}$ (REGCHAO, hereafter);  c) Those with $0.0100\ ({\rm t.u.})^{-1}\leq
L_{\rm max}$, i.e., those that are classified as chaotic for both elections of
$L_{\rm lim}$ (hereafter CHAOCHAO). Then we considered, for each orbit, eleven
$(x,y,z)$ orbital positions separated by intervals of 10 t.u., that is, over a
total interval of 100 t.u., and, for each model and for each type of orbit, we
computed the mean square value of each coordinate. Table ~\ref{regchao} gives
the square roots of the ratios of the $y$ and $z$ mean square values to the $x$
mean square value. 

\begin{table}
 \centering
  \caption{Axial ratios for different $L_{\rm lim}$. Each coordinate $x_i$
stands for $\langle x_i^2\rangle^{1/2}$.}
  \begin{tabular}{lcccc}
  \hline
   Model & ratio & REGREG & REGCHAO & CHAOCHAO \\
  \hline
   E2af & $y/x$ &
   $0.979 \pm 0.017$ & $0.987 \pm 0.018$ & $0.959 \pm 0.020$ \\
        & $z/x$ &
   $0.543 \pm 0.010$ & $0.705 \pm 0.013$ & $0.908 \pm 0.019$ \\
   E5af & $y/x$ &
   $1.069 \pm 0.024$ & $0.928 \pm 0.020$ & $0.957 \pm 0.013$ \\
        & $z/x$ &
   $0.416 \pm 0.008$ & $0.499 \pm 0.014$ & $0.556 \pm 0.007$ \\
\hline
\end{tabular}
 \label{regchao}
\end{table}

The results in Table \ref{regchao} show that the $y/x$ ratio of the E5af model
and the $z/x$ ratios of both models obtained from the REGCHAO orbits are
significantly different, at the $3\sigma$ level, from those obtained from the REGREG
orbits. We may conclude that orbits with $0.0018\ ({\rm t.u.})^{-1} \leq L_{\rm
max} < 0.0100\ ({\rm t.u.})^{-1}$ have a spatial distribution different from
that of regular orbits. Since it is in spatial distribution we are interested
here, it is thus reasonable to adopt $L_{\rm lim} = 0.0018\ ({\rm t.u.})^{-1}$. 
Therefore, we classify orbits as regular  if $L_{\rm max} < L_{\rm lim}$,  as
partially chaotic if $L_{\rm int} < L_{\rm lim} \leq L_{\rm max}$  and  as fully
chaotic if $L_{\rm lim} \leq L_{\rm int}$. The results of the classification are
shown in Table \ref{clasif}, where the quoted errors were computed as the
dispersions derived from the binomial distribution, i.e., for a percentage $p$
obtained from $N$ data, the dispersion is $\sqrt{p(100-p)/N}$.

\begin{table}
 \centering
  \caption{Percentages of chaotic and regular orbits. }
  \begin{tabular}{lccc}
  \hline
   Model & Regular & Part. chaotic & Fully chaotic\\
  \hline
   E2af & $25.80 \pm 0.70$ & $13.10 \pm 0.54$ & $61.10 \pm 0.77$ \\
   E5af & $31.93 \pm 0.80$ & $10.91 \pm 0.53$ & $57.16 \pm 0.85$ \\
\hline
\end{tabular}
 \label{clasif}
\end{table}

As in our previous investigations, chaotic orbits dominate the
dynamics of the triaxial models, with less than one third of the
bodies following regular orbits in any of them. A comparison
with the results of \cite{ZM12} suggests that chaos might be
slightly less important in the present models, but the differences
among the structures of those and the present models, from the axial
ratios onwards, makes risky any conclusion in this respect. Actually,
since rotation implies breaking a simmetry, one might expect to
find more chaos in rotating models, as indicated by \cite{M06}
who, in fact, found slightly more chaos in a very slowly rotating
model than in the same model without rotation. 

The long integration interval used to obtain the FT-LCNs allows us to make a
further check of the possible escapes searching for those bodies that end up
farther from the center of the system than the farthest body in the original
model. We found that, after 10,000 t.u., only 1 body (i.e. 0.03
per cent) on a fully 
chaotic orbit escapes from the E2af model, while 49 bodies (1.42
per cent) on 
partially chaotic orbits and 80 bodies (2.33 per cent) on
fully chaotic orbits escape 
from the E5af model, according to this criterium. Two remarks should be made on
this respect. First, there is no guarantee that those bodies actually escape
because, on the one hand, many of them do not exceed the adopted distance limit
by a significant amount and, on the other hand, other bodies alternate shorter
and longer distances at intermediate times, i.e., they seem to be on highly
elongated rather than on open orbits. Second, strictly speaking, a chaotic orbit
should have a positive FT-LCN \emph{and} to be bound (e.g., orbits in a
repulsive harmonic potential are perfectly regular, in spite of diverging
exponentially, but they are unbound), so that the percentages of chaotic orbits
of Table \ref{clasif} might include fractions of unbound orbits that are not
chaotic. Nevertheless, those fractions should be exceedingly small in view of
our results on the final distances after 10,000 t.u. Moreover, considering that
that interval amounts to about 40 Hubble times, we may conclude that escapes
are not a serious problem for our models and that they are very stable indeed.

\begin{figure*}
\resizebox{\hsize}{!}{\includegraphics{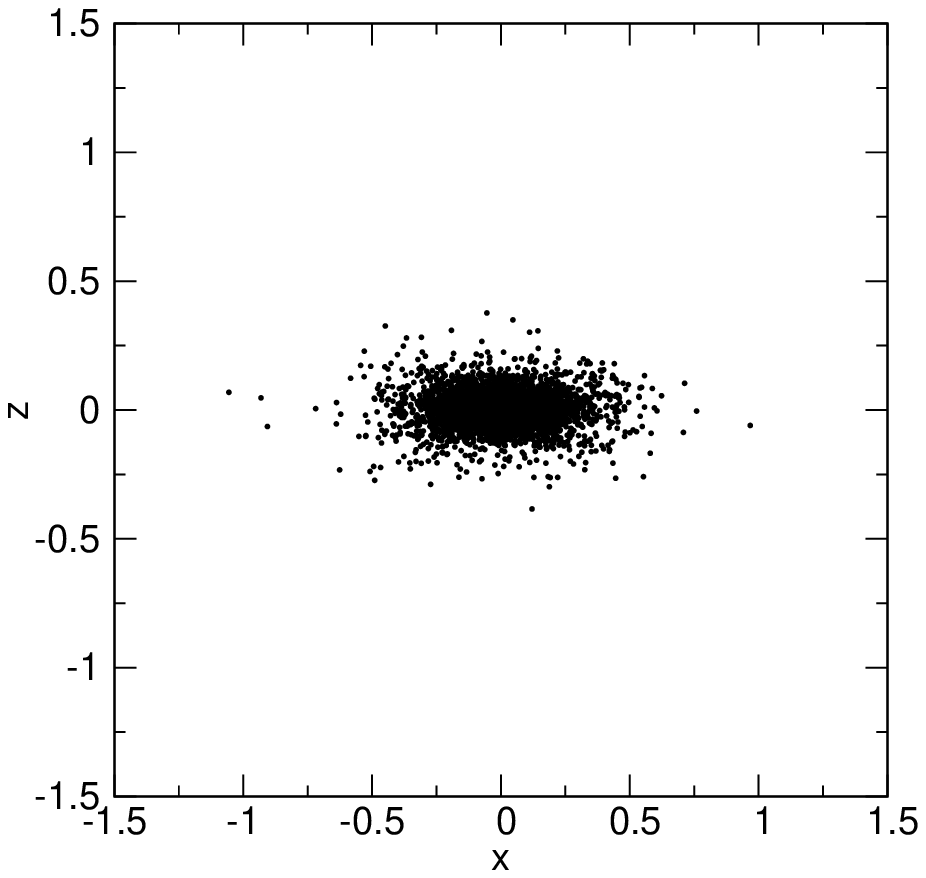}\hspace{1cm}
                      \includegraphics{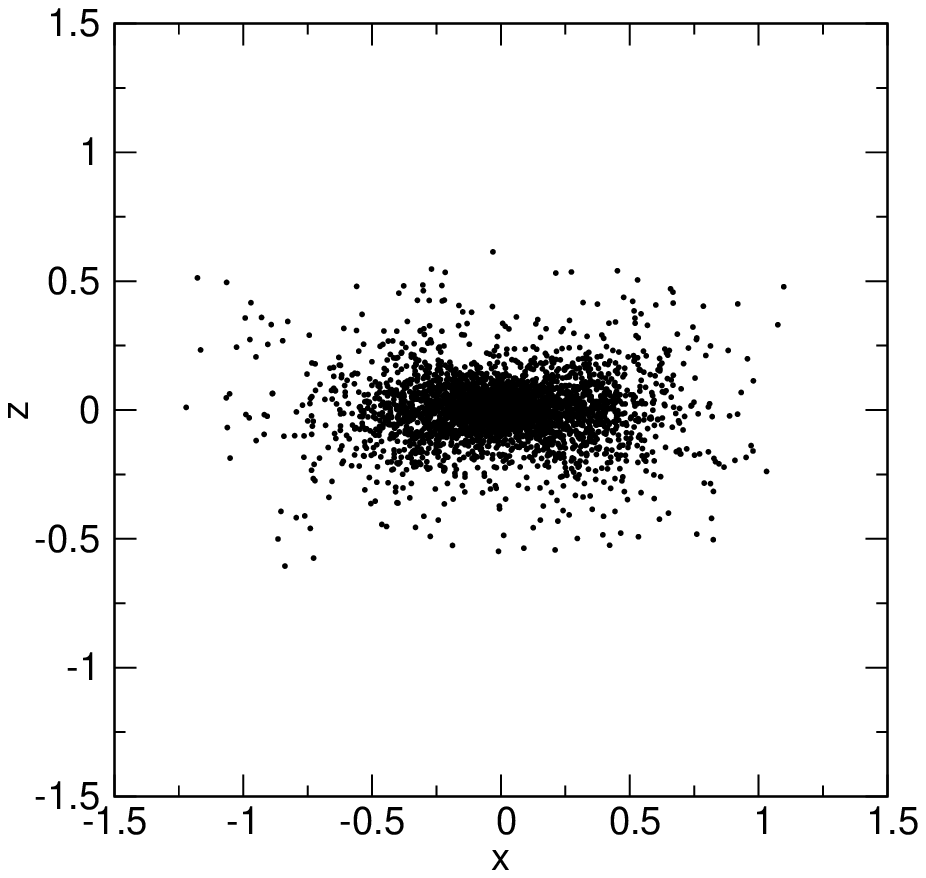}\hspace{1cm}
                      \includegraphics{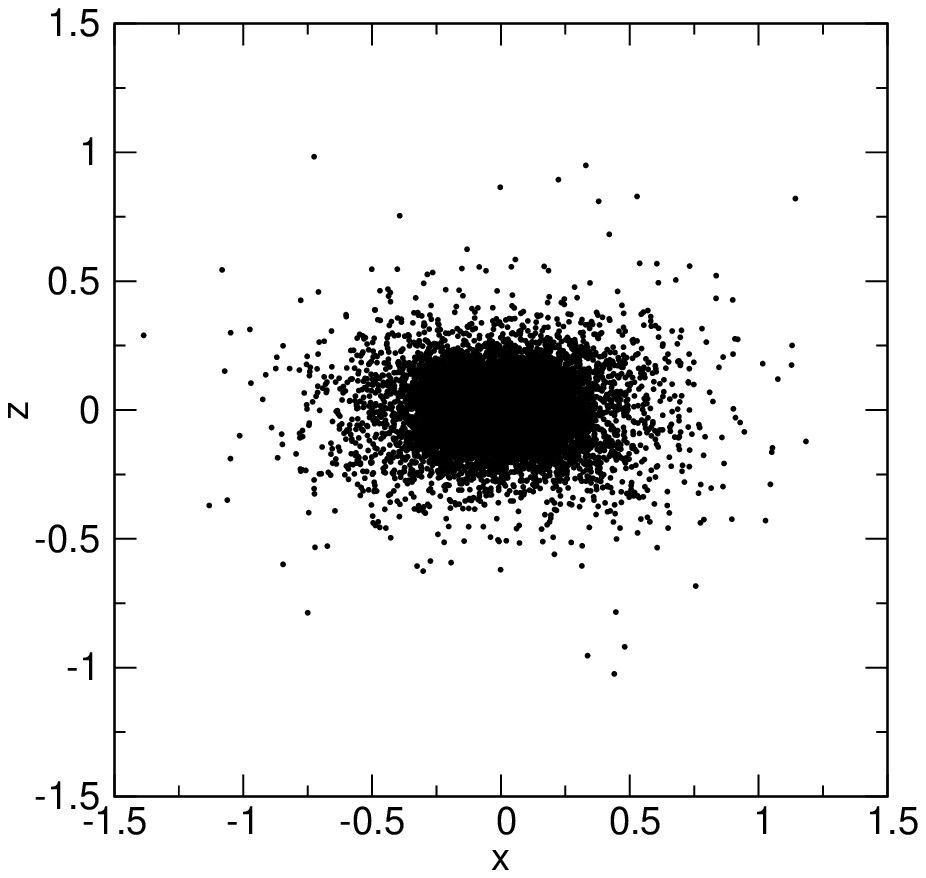}}
\resizebox{\hsize}{!}{\includegraphics{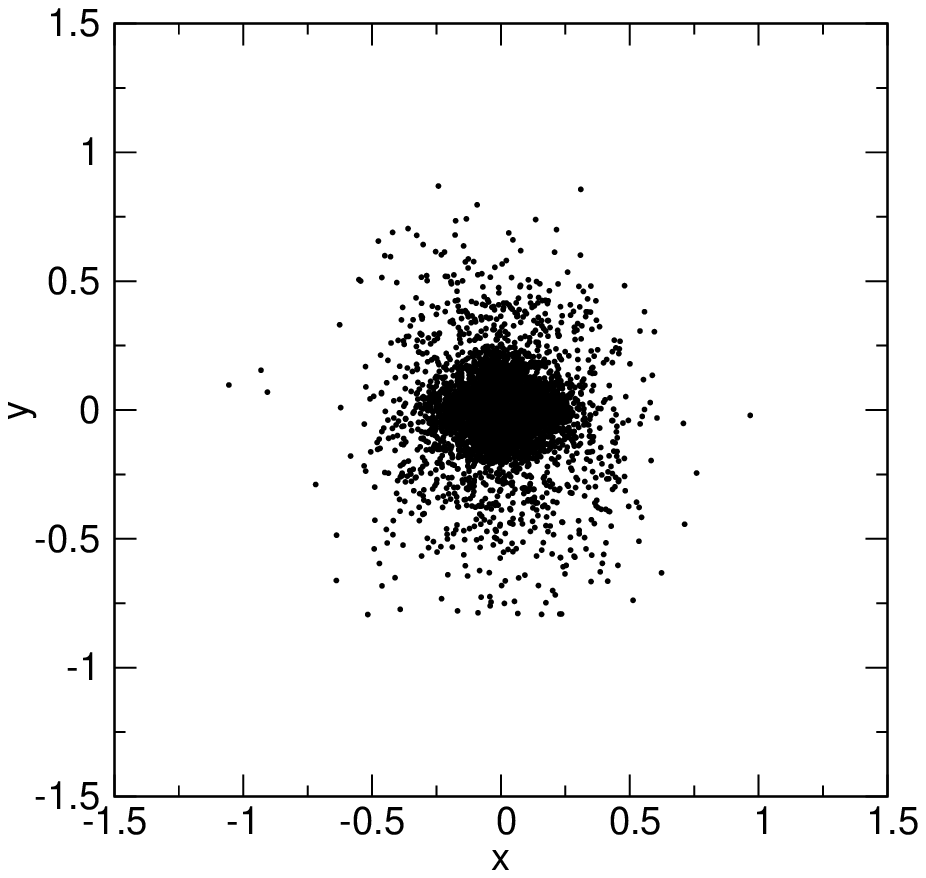}\hspace{1cm}
                      \includegraphics{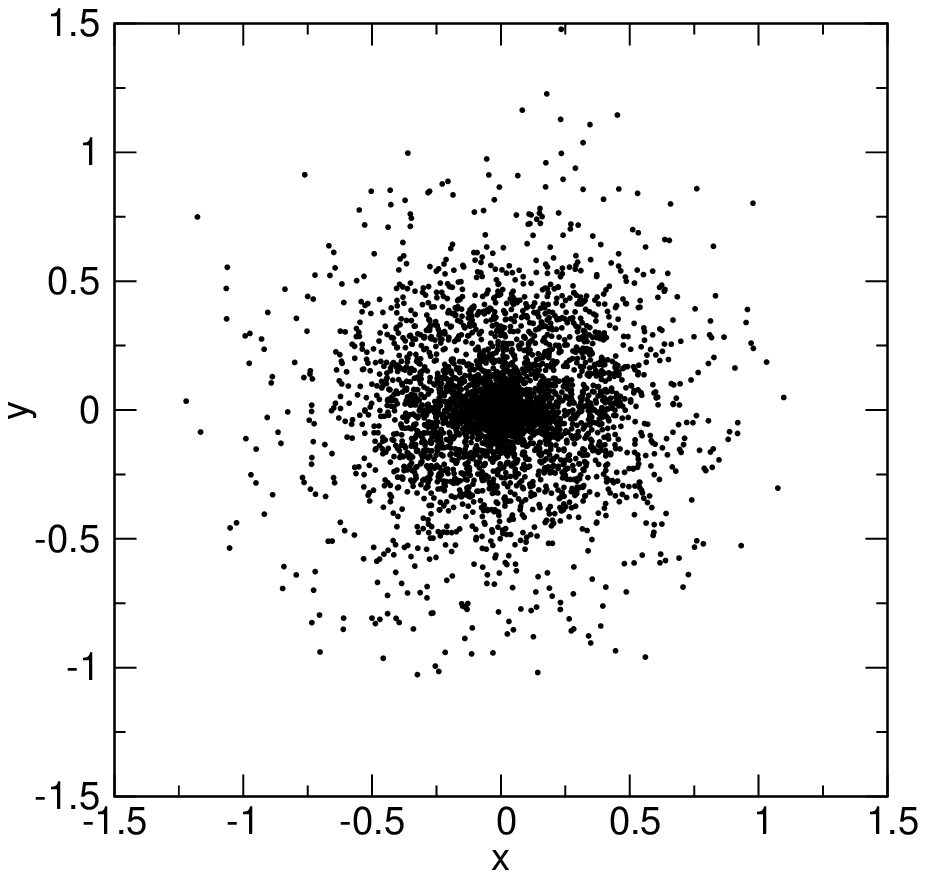}\hspace{1cm}
                      \includegraphics{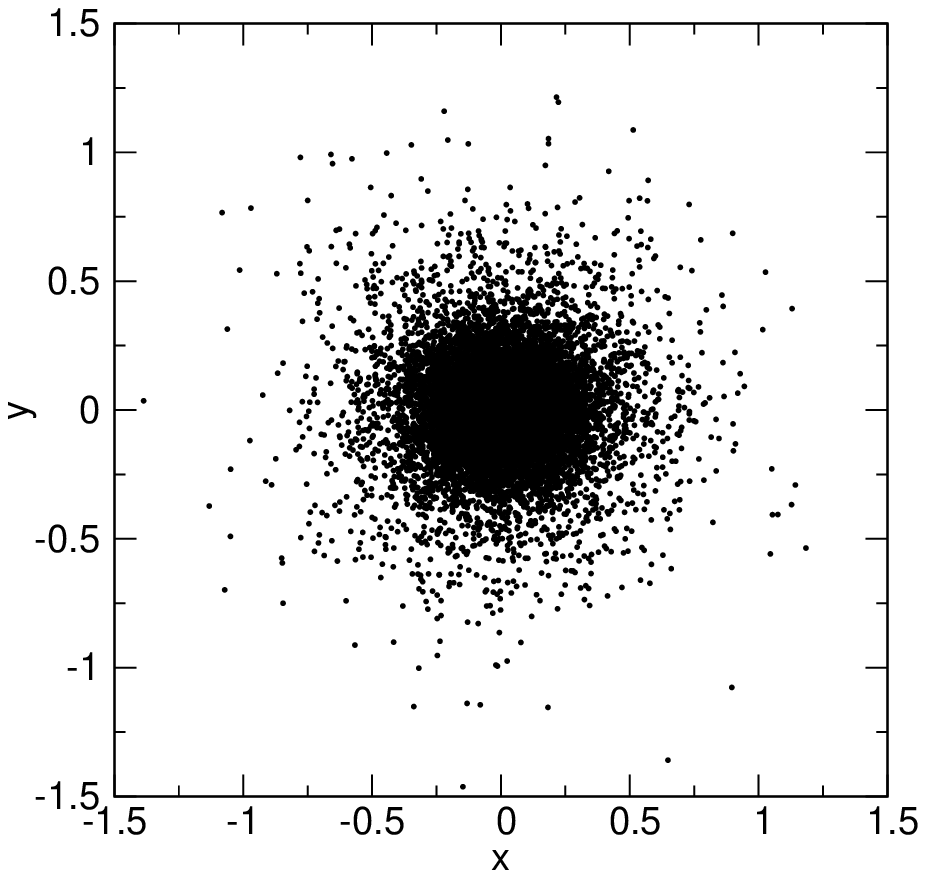}}
\caption{Distribution of the regular (left), partially (center) and fully chaotic
(right) orbits for model E5af.} 
\label{xyzrpf}
\end{figure*}

Fig. \ref{xyzrpf} presents the $(x, y)$ and $(x, z)$ projections of the bodies
in the E5af model whose FT-LCN had been computed, separately for those in
regular, partially and totally chaotic orbits. To increase the number of points,
we have plotted the positions at 10 t.u. intervals over a total integration time
of 100 t.u., i.e., 11 points for each orbit. As in our previous investigations
on triaxial models, the distribution of the fully chaotic orbits is very
different from that of the regular ones, while the partially chaotic orbits
adopt a distribution intermediate between the other two. This visual impresion
is confirmed by the results of Table \ref{axrpf} where we present, for the two
models and separately for the regular, partially and fully chaotic orbits, the
values of the axial ratios computed from the positions at 10 t.u. intervals from
the initial 100 t.u. of the integration interval, i.e., like in Fig.
\ref{xyzrpf}. The very different distributions of the regular and fully chaotic
orbits is most clearly shown by their different $z/x$ values, although the
difference between the corresponding $y/x$ values of the E5af model is also
significant at the $3\sigma$ level. Partially chaotic orbits have $z/x$ values
intermediate between those of the regular and fully chaotic orbits in both
models. For the E2af model there are no significant differences, at
the $3\sigma$ level, among the $y/x$ ratios, but there are very significant
ones among the $z/x$ values. At any rate, it
is clear that the distributions of the regular, partially and fully chaotic
orbits are very different from each other.

\begin{table}
 \centering
  \caption{Axial ratios of regular, partially and fully chaotic orbits. Each
coordinate $x_i$ stands for $\langle x_i^2\rangle^{1/2}$.}
  \begin{tabular}{lcccc}
  \hline
   Model & ratio & Regular & Part. chaotic & Fully chaotic\\
  \hline
   E2af & $y/x$ & $0.979 \pm 0.017$ & $0.994 \pm 0.018$ & $0.942 \pm 0.018$ \\
        & $z/x$ & $0.543 \pm 0.010$ & $0.679 \pm 0.012$ & $0.967 \pm 0.019$ \\
   E5af & $y/x$ & $1.069 \pm 0.024$ & $0.948 \pm 0.021$ & $0.956 \pm 0.013$ \\
        & $z/x$ & $0.416 \pm 0.008$ & $0.424 \pm 0.010$ & $0.602 \pm 0.008$ \\
\hline
\end{tabular}
 \label{axrpf}
\end{table}

\subsection{Regular orbits}

The frequency analysis was performed, as in our previous papers \citep{M06,
AMNZ07, MNZ09,MNZ13,CMN14}, with the modified Fourier transform code of
\citet{SN96} (a copy can be obtained at www.boulder.swri.edu/$\sim$davidn) and 
adopting as initial conditions the positions and velocities of the same bodies
we had selected for the computation of the Lyapunov exponents. For each one of
the 2,117 regular orbits of the two models, we obtained the fundamental
frequencies for each coordinate, $F_x$, $F_y$ and $F_z$, through the frequency
analysis of the complex variables $x+{\rm i}\,\dot x$, $y+{\rm i}\,\dot y$ and
$z+{\rm i}\,\dot z$, respectively; these were derived from 8,192 points equally
spaced in time obtained integrating the orbits over 300 radial periods. As
indicated by \citet{M06}, the frequencies of isolated lines obtained in this way
have errors smaller than $10^{-9}$, but the precision is much lower when there
are nearby lines, and here we adopt the practical limit of $2 \times 10^{-4}$
for the precision, as in our previous works.

We obtained the fundamental frequencies using the method of \citet{KV05} with the
improvements introduced by \citet{M06}, \citet{AMNZ07} and \citet{MNZ09}.
The original method took the frequency of the largest amplitude in each
coordinate as the fundamental frequency for that coordinate but, as shown by
\citet{BS82} and \citet{M06}, respectively, the libration of some orbits and
the extreme elongation of others makes necessary to adopt, for these cases, other
frequencies as the fundamental ones, and the improvements deal with those cases.

In our previous work on non-, or very slowly, rotating systems, the fundamental
frequencies were used to classify the regular orbits as long and short axis
tubes (LATs and SATs, respectively) and boxes and boxlets (BBLs), but the orbital
composition of rotating systems is much more varied and complex \citep[see, e.g.,]
[]{BT08} and that classification is not enough. Besides, although it is not
unfrequent to find those same names used for orbits in rotating systems, it should
be recalled that things are quite different in fixed and rotating systems of
reference (e.g., a BBL in one system may be a SAT in another one). As we
will see below, most of the regular orbits in our models are resonant, i.e.,
their fundamental frequencies obey one or two equations of the form:

\begin{equation}
l F_x + m F_y + n F_z = 0,
\end{equation}
with $l, m$ and $n$ integers not all equal to zero. Therefore, in what follows
we will refer to the different orbits mainly by their resonances, i.e., $(l, m, n)$
in the equation above. Besides, we will not restrict ourselves to frequency
analysis and we will add other criteria (like the conservation of the
signs of the components of the angular momentum and plots of the orbits) to aid
the orbital classification.

\begin{figure*}
\resizebox{\hsize}{!}{\includegraphics{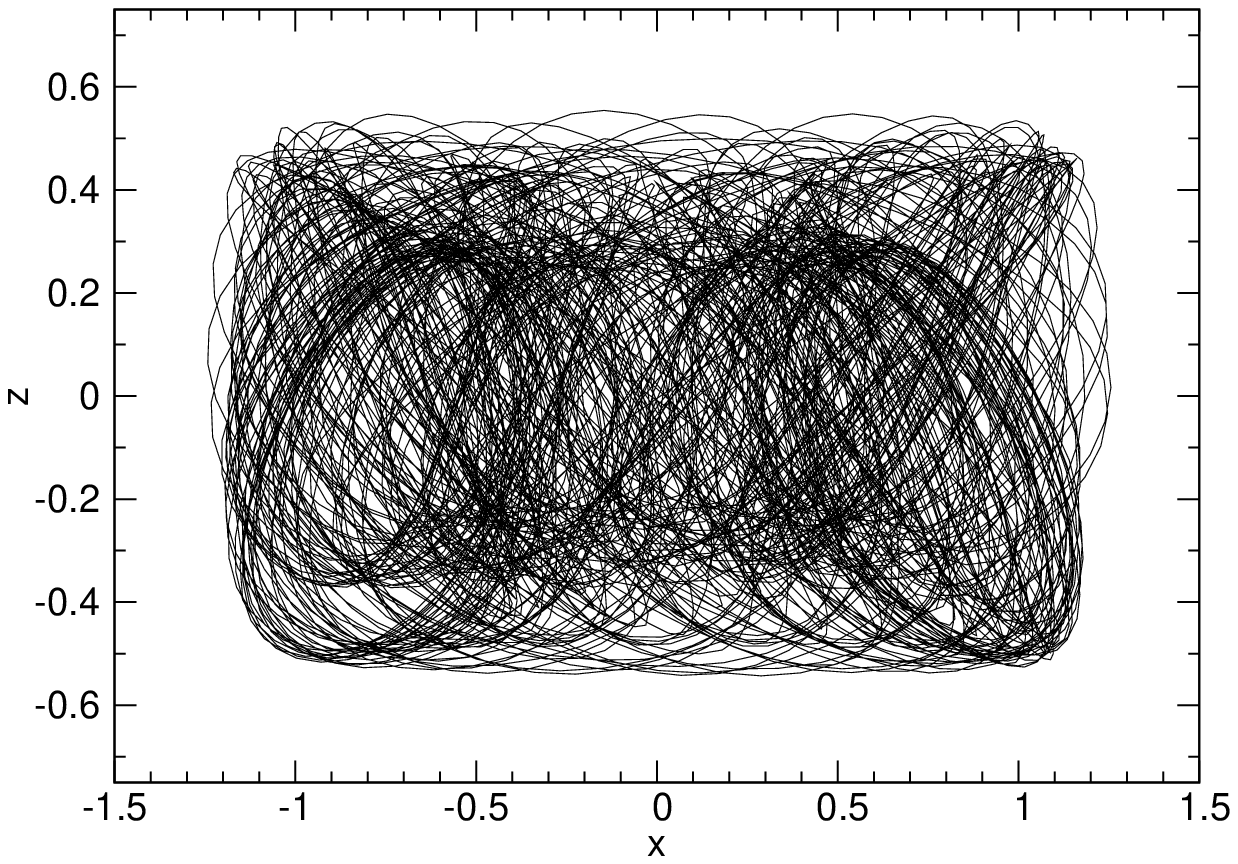}\hspace{0cm}
                      \includegraphics{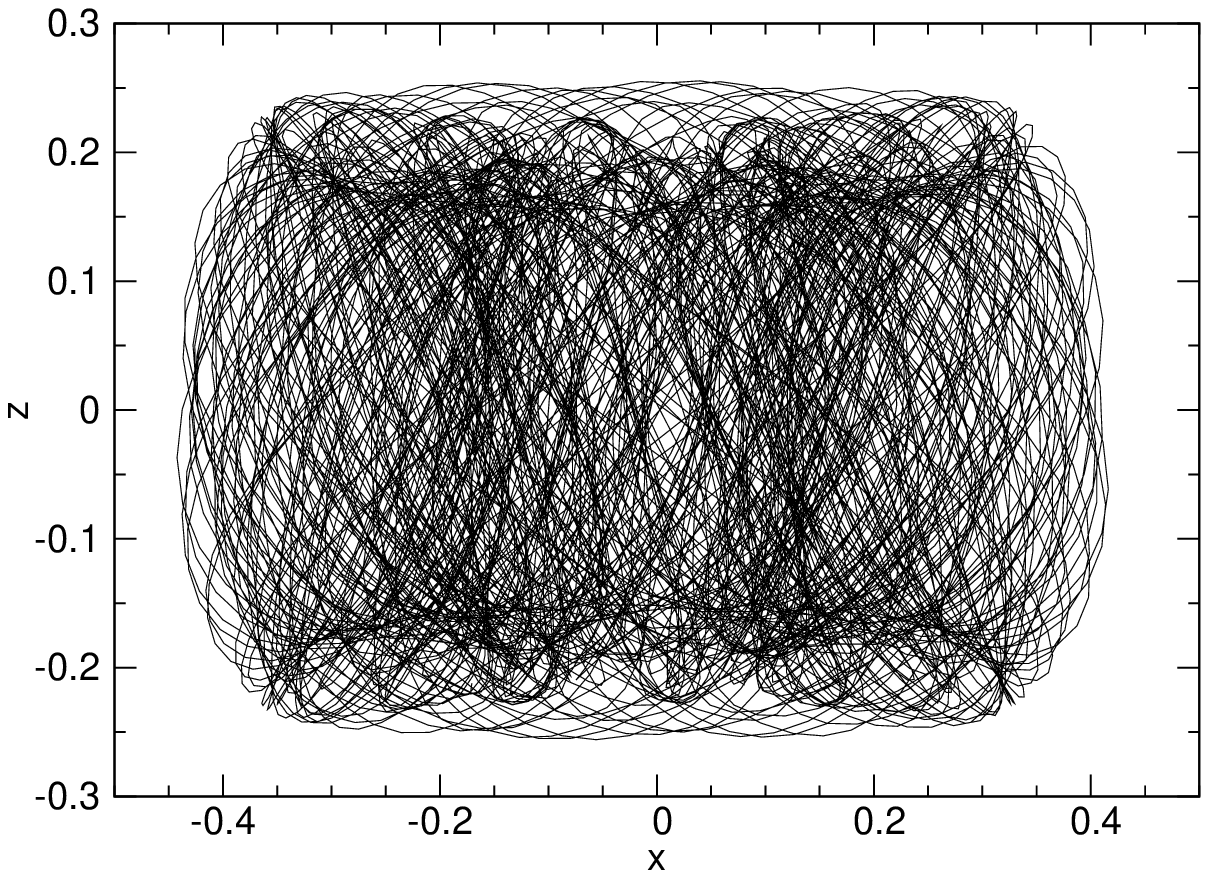}}
\resizebox{\hsize}{!}{\includegraphics{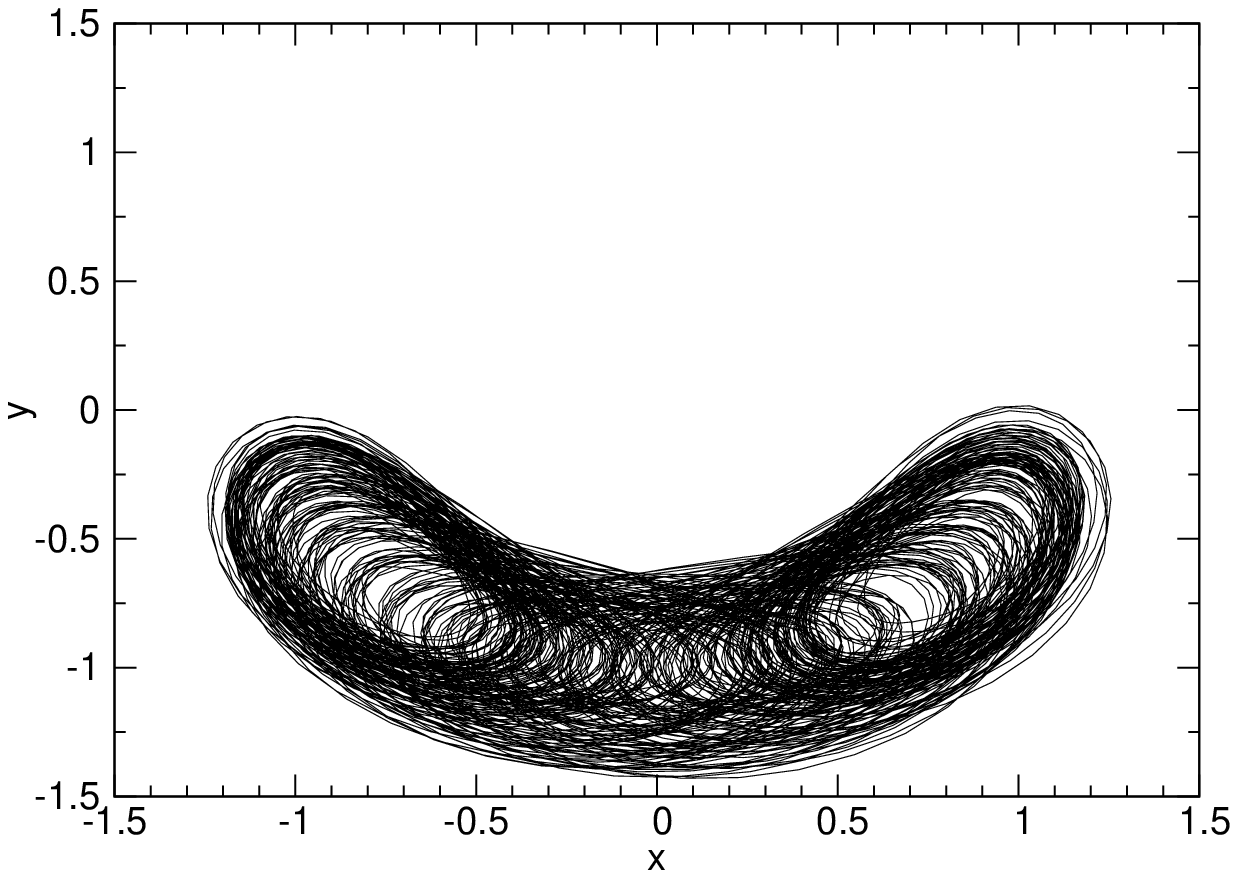}\hspace{0cm}
                      \includegraphics{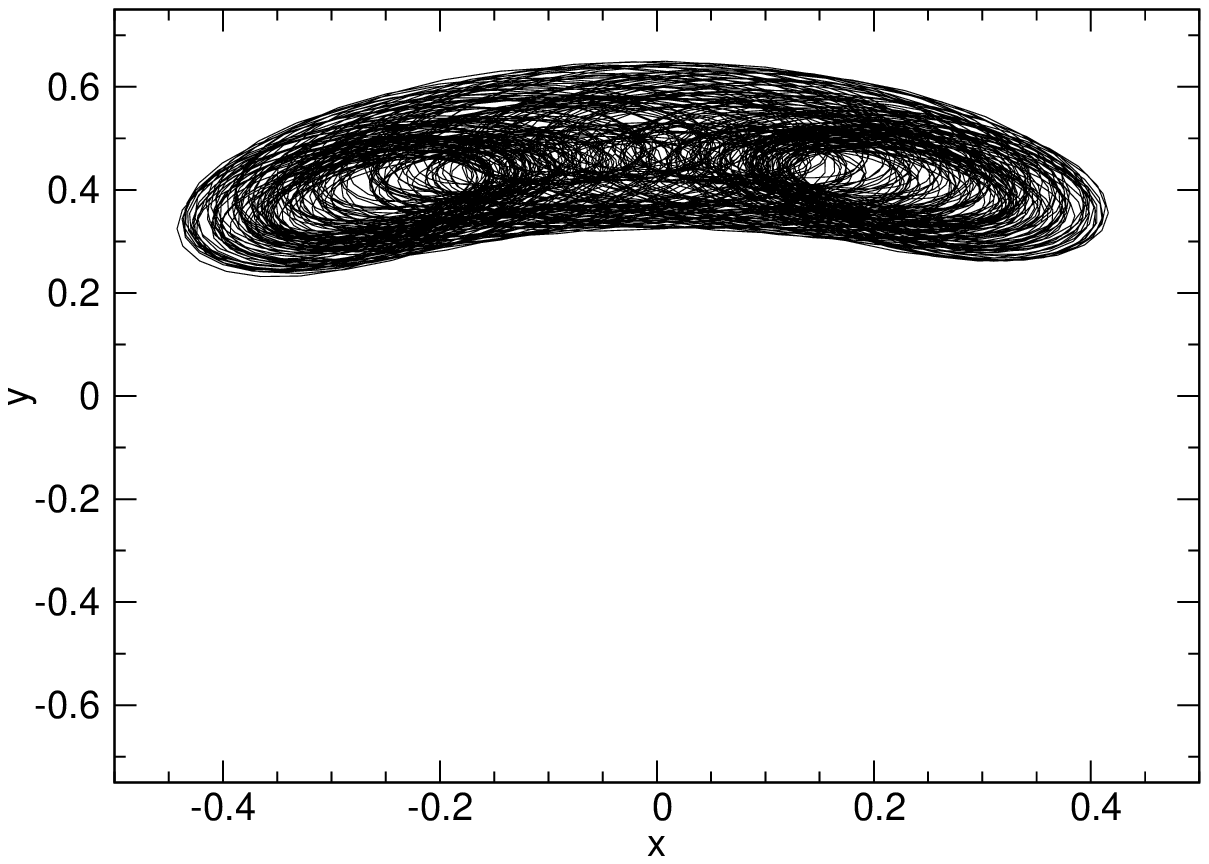}}
\caption{Horseshoe orbits 231 of model E2af (left) and 182 of model E5af
(right).} 
\label{shoe}
\end{figure*}

Of the 2,117 orbits regarded as regular, 122 yielded values of their fundamental
frequencies that did not obey that $F_x \le F_y \le F_z$. As in our previous
works, visual inspection of their spectra showed that many of them were typical
of chaotic orbits, with lines of similar frequencies and amplitudes. We checked
that possibility obtaining the FT-LCNs of those orbits using an integration time
of 100,000 t.u., i.e. ten times longer than the one of Subsection \ref{subchao},
and 44 of the 122 suspicious orbits turned out to be actually chaotic. Besides,
it should be recalled that in rotating systems there are orbits whose fundamental
frequencies might not be obtained with the same criteria used for non--rotating
ones. For example, another 11 of those 78 orbits had a frequency line with
$F_y$ close to zero, i.e., they were horseshoe orbits (two examples are shown
in Fig. \ref{shoe}).

\begin{figure*}
\resizebox{\hsize}{!}{\includegraphics{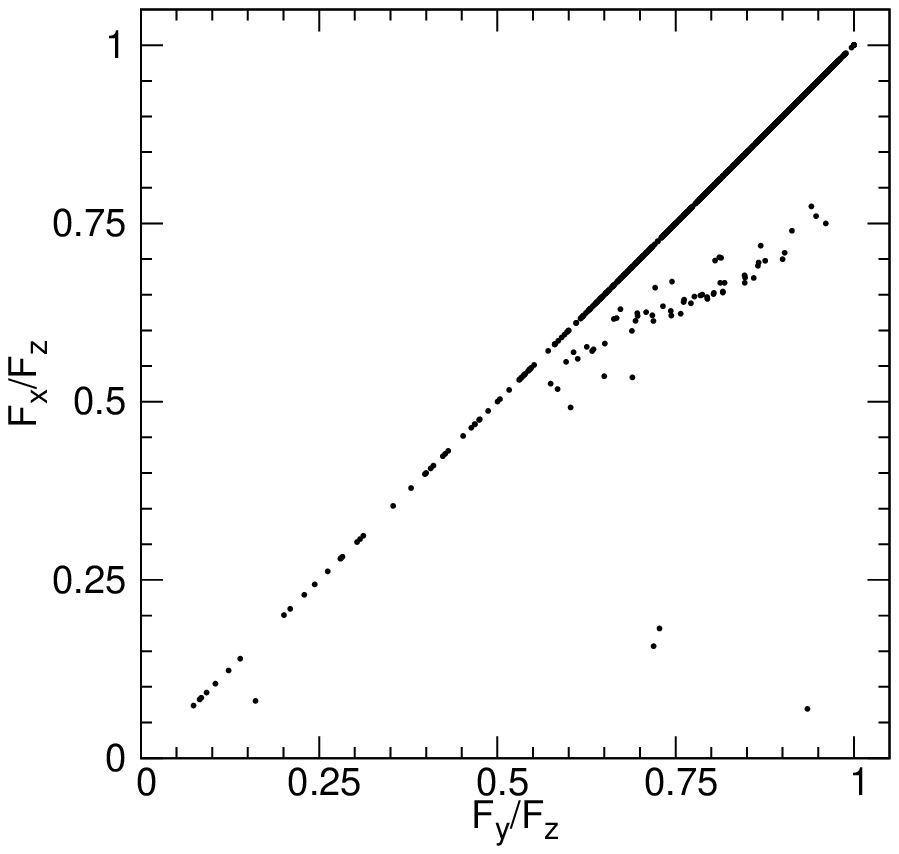}\hspace{1cm}
                      \includegraphics{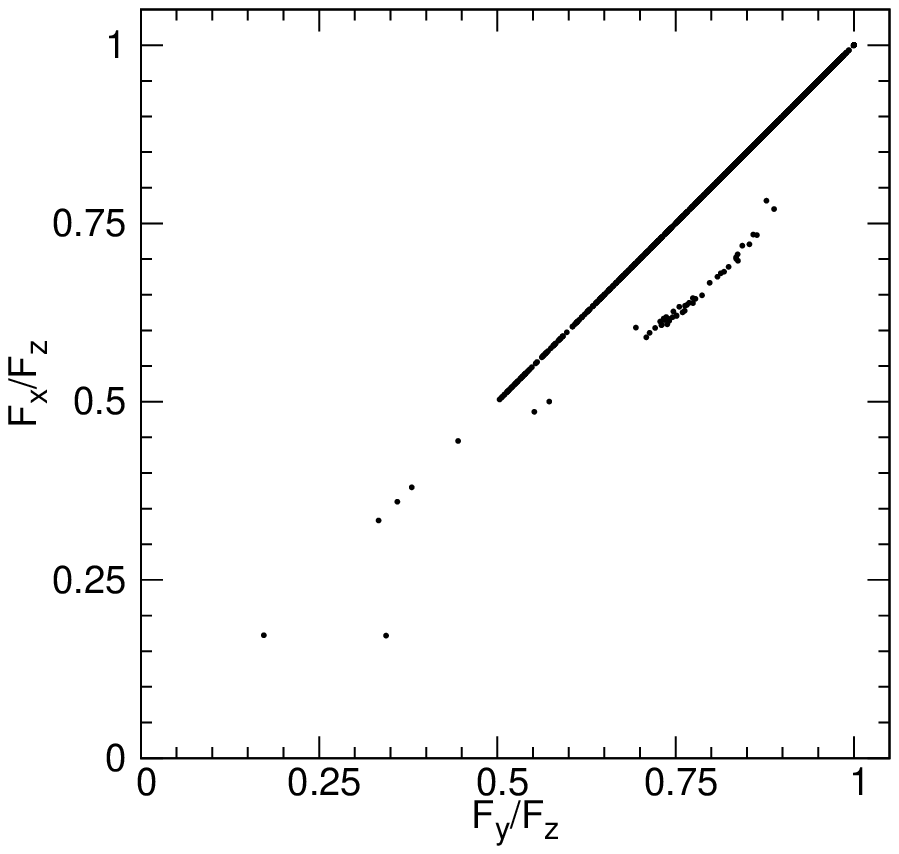}}
\caption{Left: Frequency map of the E2af model. Right: Frequency map of the E5af
model.}
\label{fmap}
\end{figure*}

Fig. \ref{fmap} presents the frequency maps for both models. Orbits that do
not obey $F_x \le F_y \le F_z$, mentioned above, are not included, but we notice
four cases of model E2af and one of model E5af in odd positions of the diagrams.
Orbit 2166 of model E2af, at $(F_y/F_z,F_x/F_z)=(0.161, 0.080)$, and orbit 773
of model E5af, at $(0.344, 0.172)$, are just horseshoes. Visual inspection of
the frequency spectra of orbits 1706, at $(0.727, 0.182)$, and 1870, at $(0.719,
0.157)$, of model E2af showed that they are very complex, making very difficult
to select the fundamental frequencies. Plots of those orbits suggest that they
are boxlets that avoid the center of the system; besides, they do not conserve
the sign of the components of the angular momentum and both obey the $(1, -3,
2)$ resonance. The frequency spectrum of orbit 3390 of model E2af, at $(0.935,
0.069)$, was clearly chaotic and the chaoticity of that orbit was confirmed
recomputing its FT-LCNs with a 100,000 u.t. interval.

\begin{figure*}
\resizebox{\hsize}{!}{\includegraphics{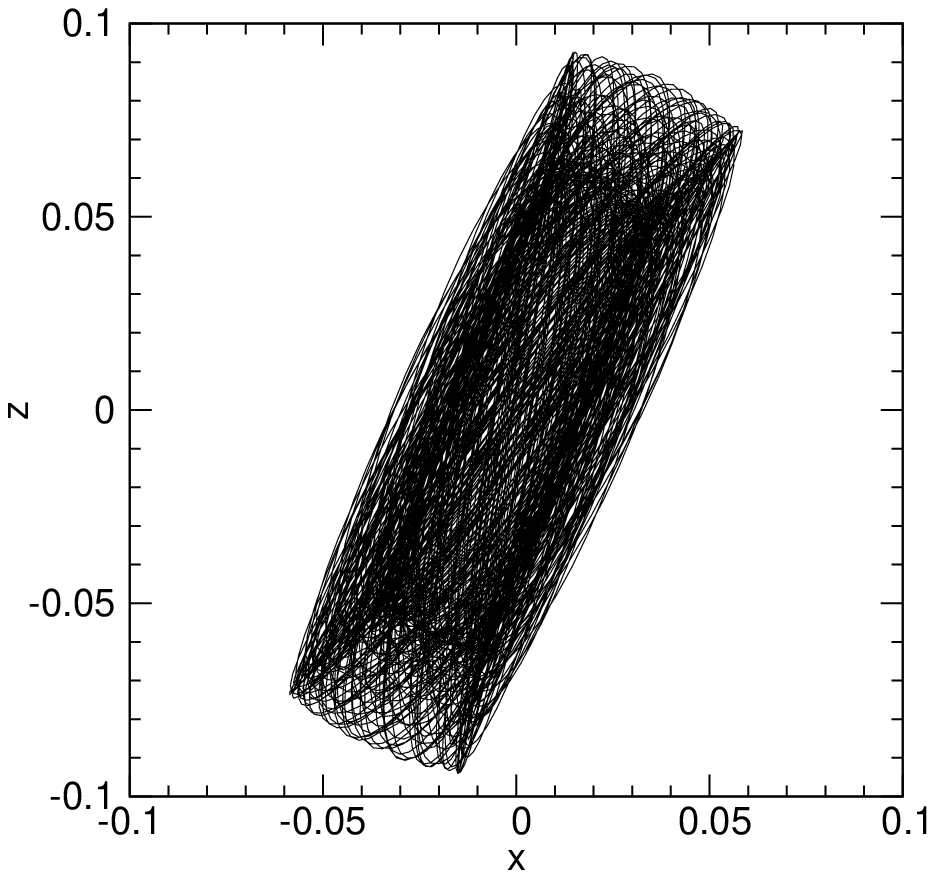}\hspace{1cm}
                      \includegraphics{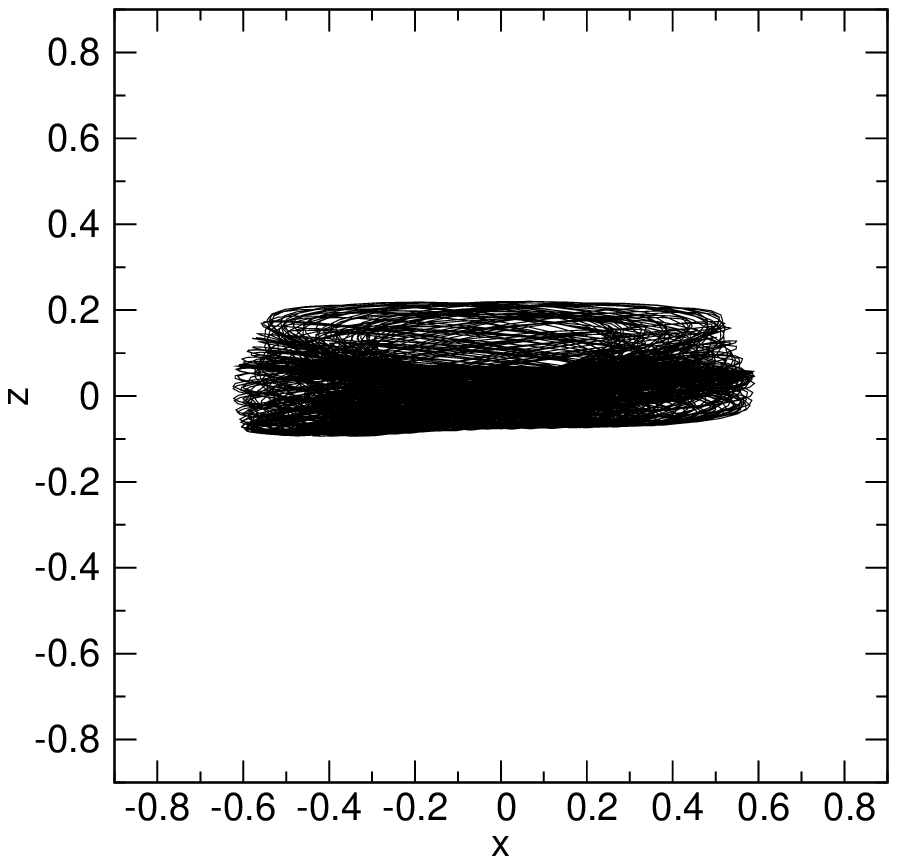}}
\resizebox{\hsize}{!}{\includegraphics{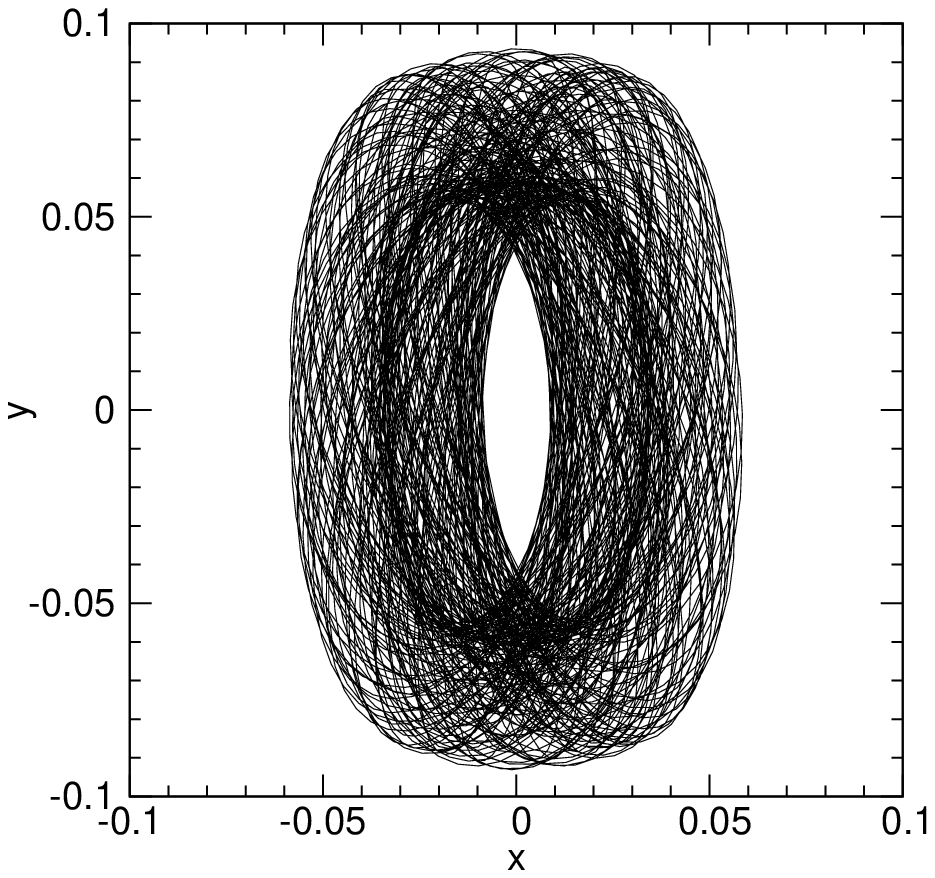}\hspace{1cm}
                      \includegraphics{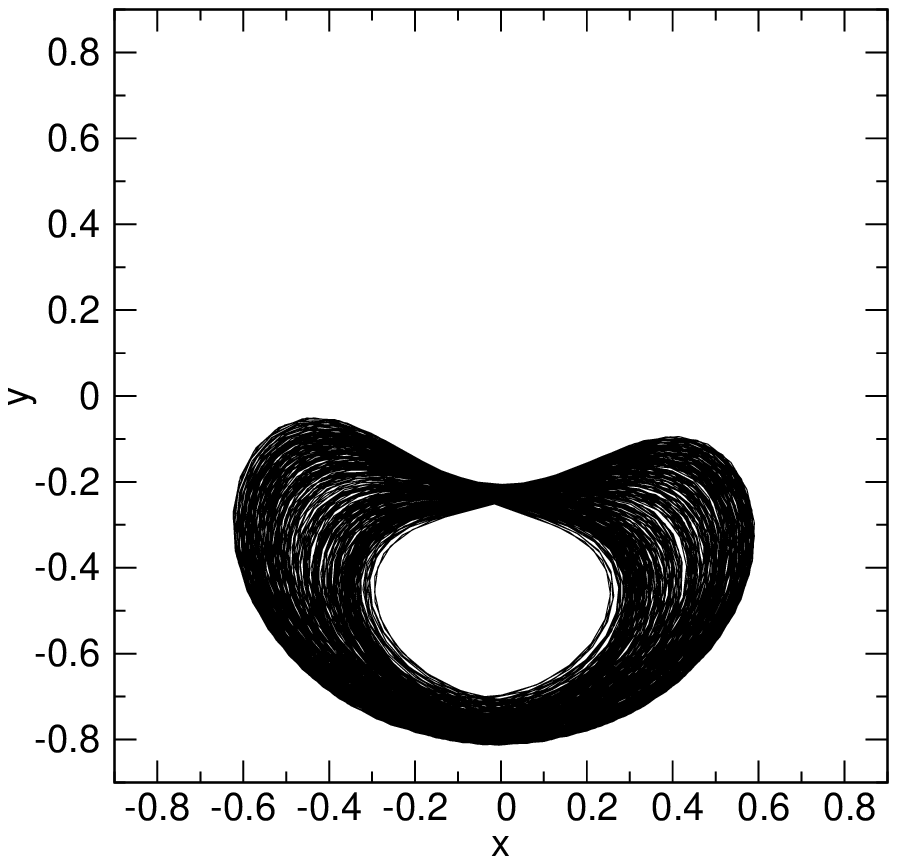}}
\caption{Tilted tube orbit 521 of model E2af (left) and closed horseshoe orbit
177 of model E5af (right).} 
\label{satlat}
\end{figure*}

The most striking feature of Fig. \ref{fmap} is the complete lack of
representative points at $F_y/F_z \simeq 1.0$ for $F_x/F_z < 1.0$, i.e.,
the $(0, 1, -1)$ resonant orbits, or the LATs of non--rotating systems.
Nevertheless, the points at  $F_y/F_z \simeq 1.0$ and $F_x/F_z \simeq 1.0$
are not single points but each one includes many orbits (35 for model
E2af and 61 for model E5af) that obey both the $(1, -1, 0)$ and
$(0, 1, -1)$ resonances.
All those from the E2af model conserve the sign of the $x$ component
of the angular momentum, but only 13 of them conserve the sign of
the $z$ component as well: they are tubes with their axes in the
$(x, z)$ plane, but not in the $(x, y)$ plane, i.e., they are inclined
with respect to the latter plane, and the conservation of the sign of
the $z$ component depends on the tilt and the width of the tube. An
example is shown on the left side of Fig. \ref{satlat}. 59 of the 61
cases of the E5af model are also tilted tubes like those of
model E2af, but their tilts and widths are such that all conserve
the signs of both the $x$ and $z$ components of the angular momentum.
The remaining two cases do not conserve the sign of
the $x$ component of the angular momentum and plots of those orbits
showed that they are tubes around the Lagrangian point L5, or like
horseshoes long enough to close onto themselves. In fact, when the
angular momentum is computed with respect to L5, the sign of
its $z$ component is conserved for those orbits. One of them is
shown on the right side of Fig. \ref{satlat}.

\begin{figure*}
\resizebox{\hsize}{!}{\includegraphics{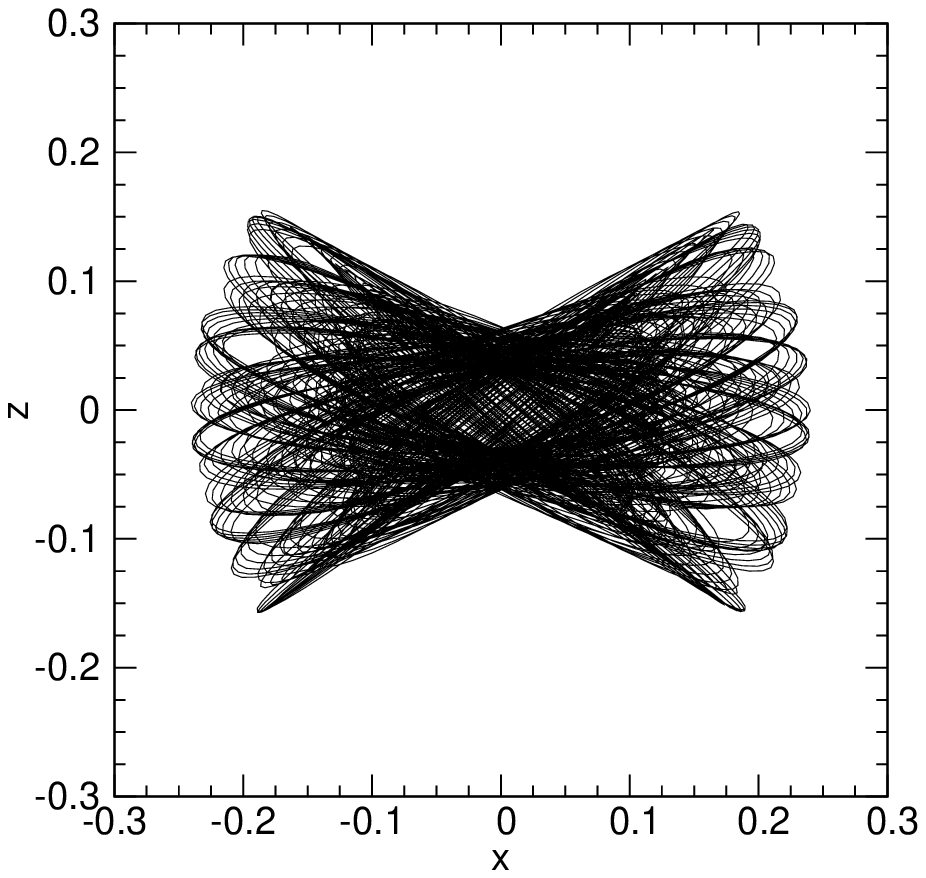}\hspace{1cm}
                      \includegraphics{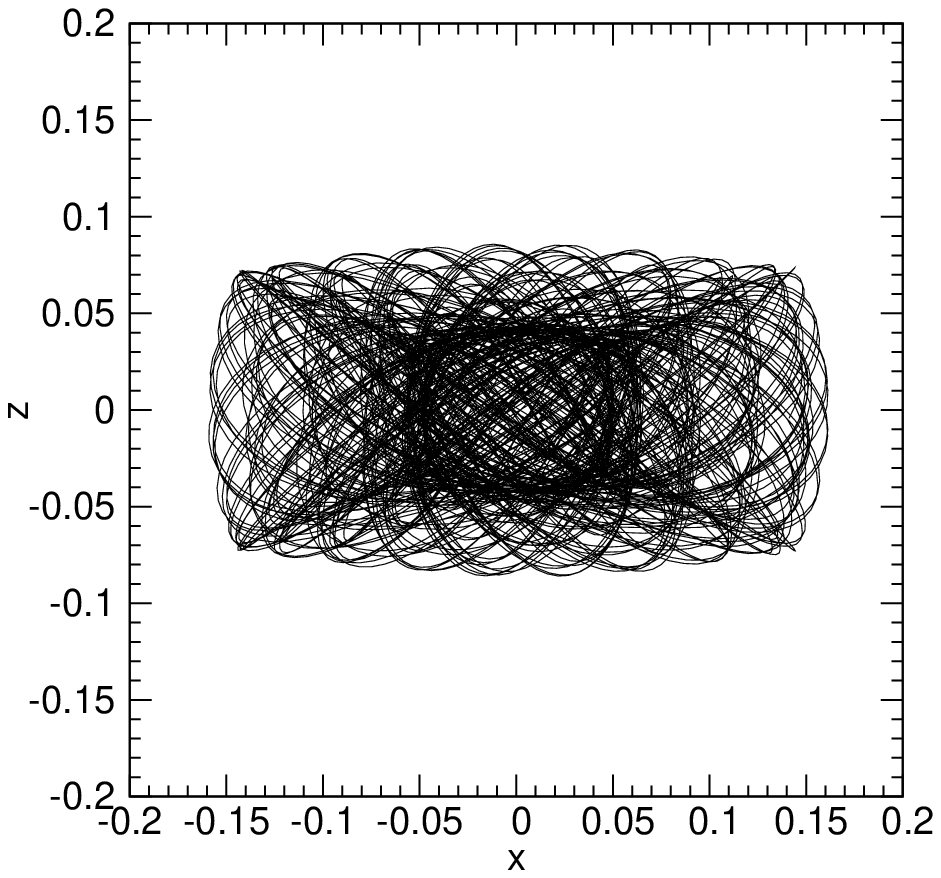}}
\resizebox{\hsize}{!}{\includegraphics{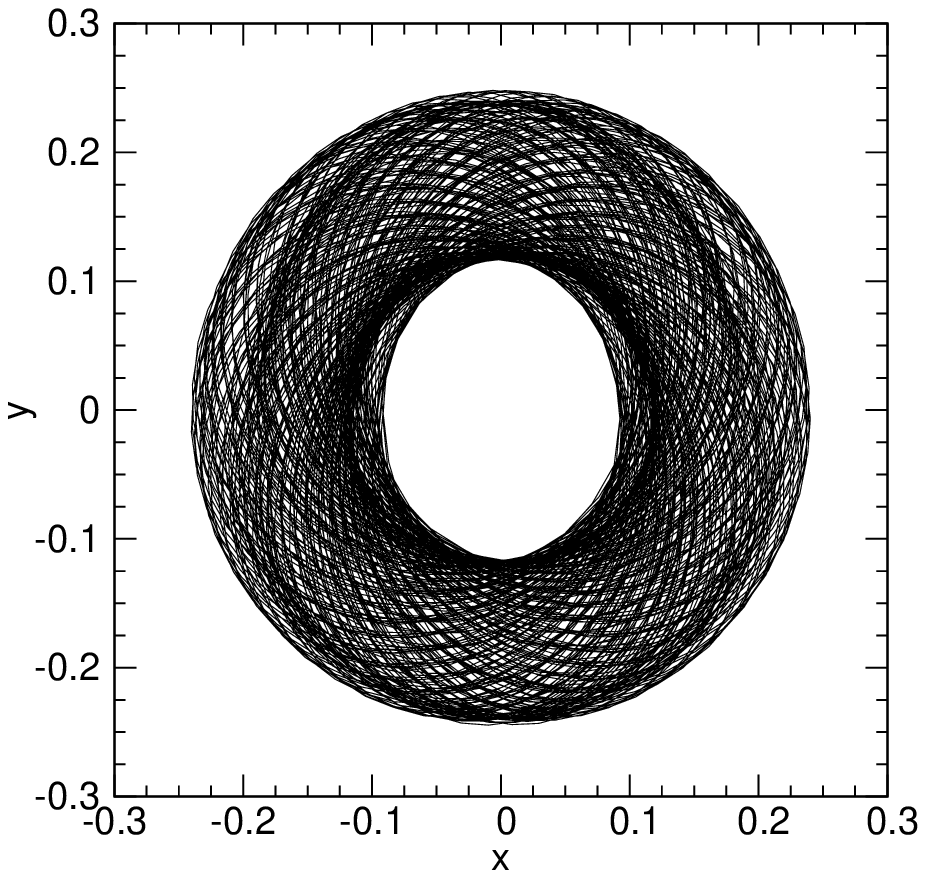}\hspace{1cm}
                      \includegraphics{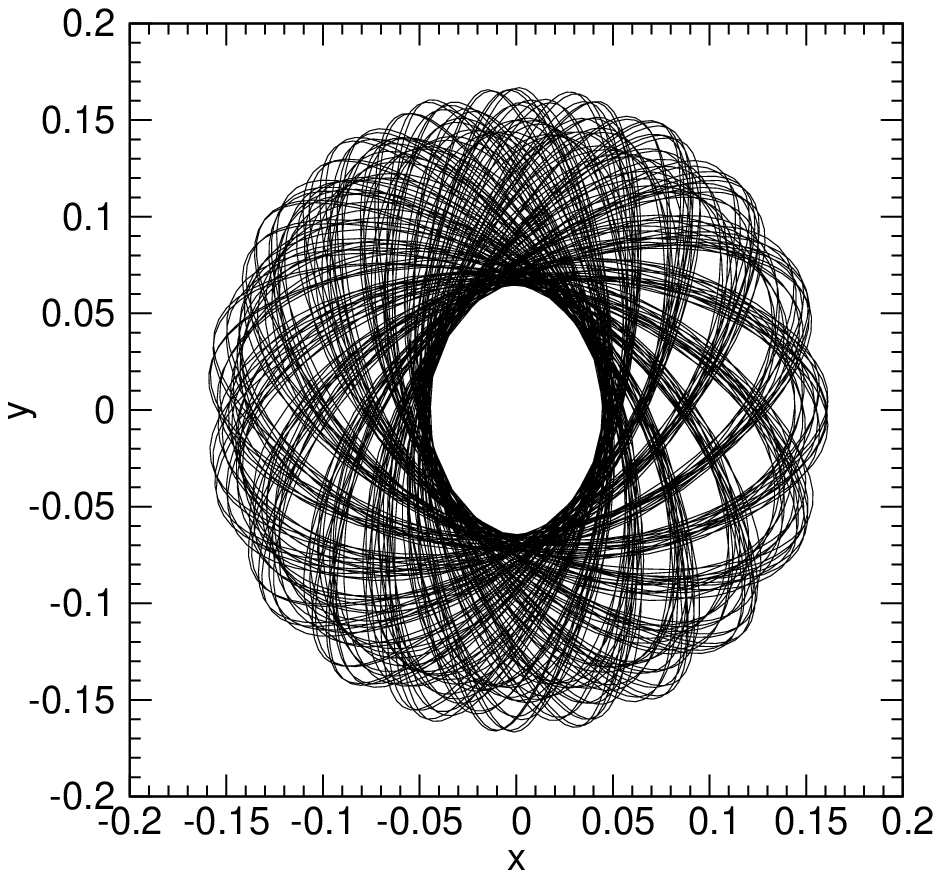}}
\caption{Two SATs: orbits 76 of model E2af (left) and 20 of model E5af (right).} 
\label{sats}
\end{figure*}

\begin{figure*}
\resizebox{\hsize}{!}{\includegraphics{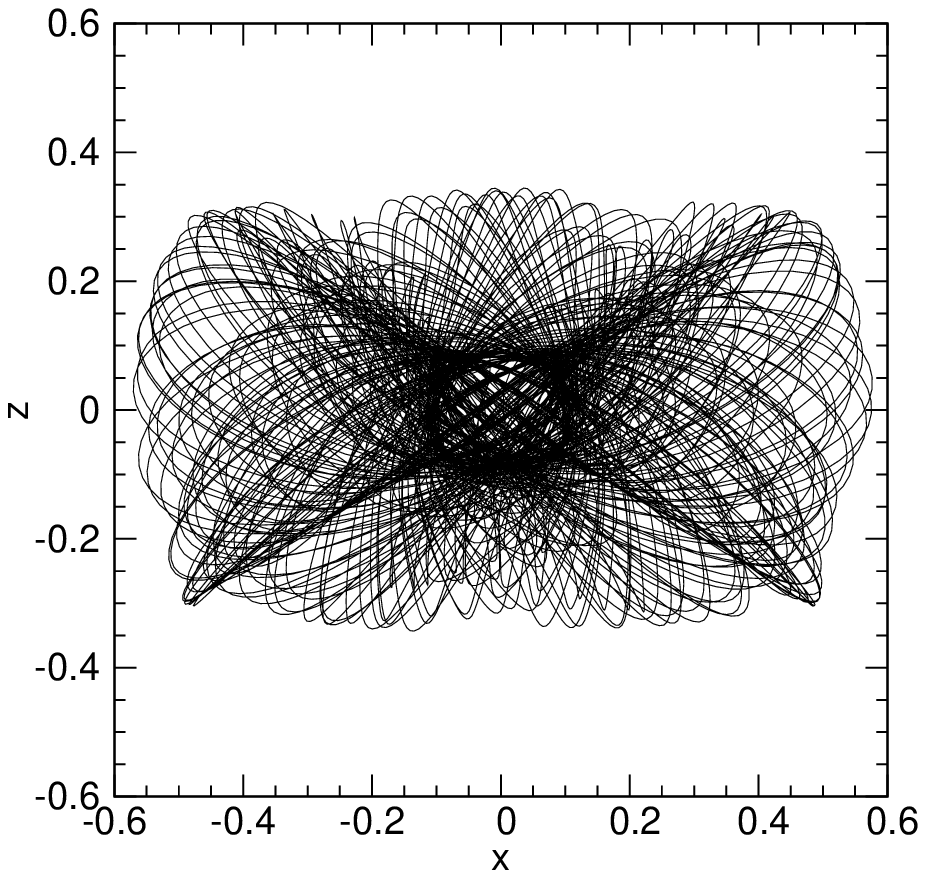}\hspace{1cm}
                      \includegraphics{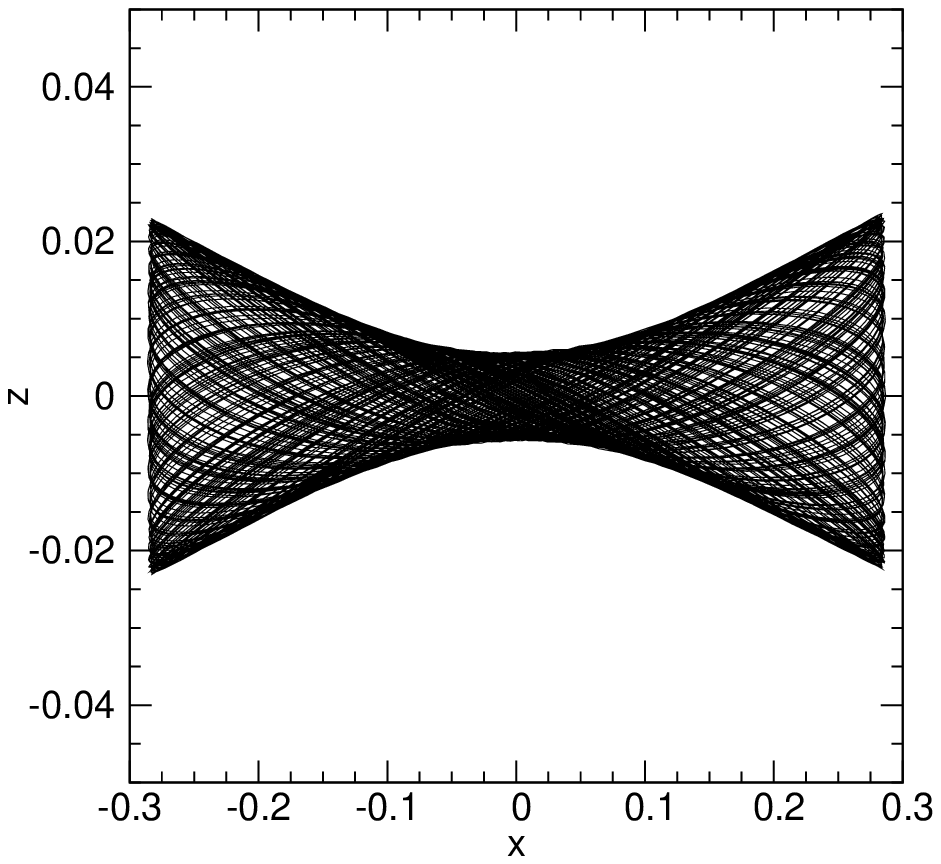}}
\resizebox{\hsize}{!}{\includegraphics{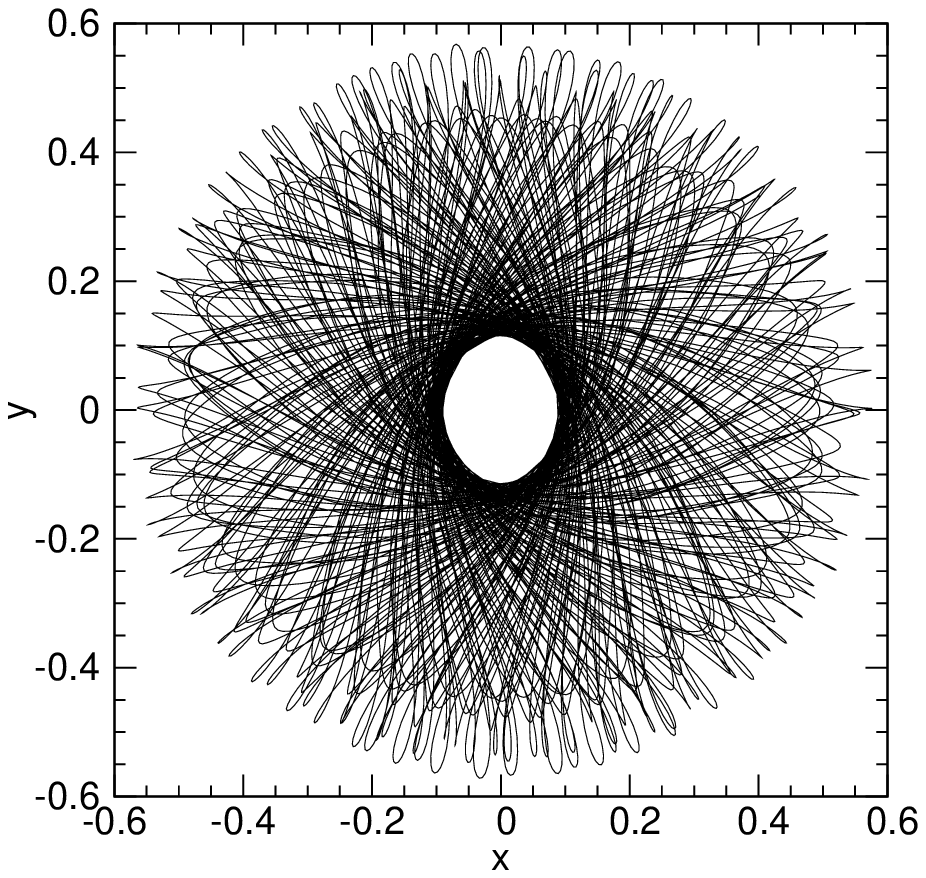}\hspace{1cm}
                      \includegraphics{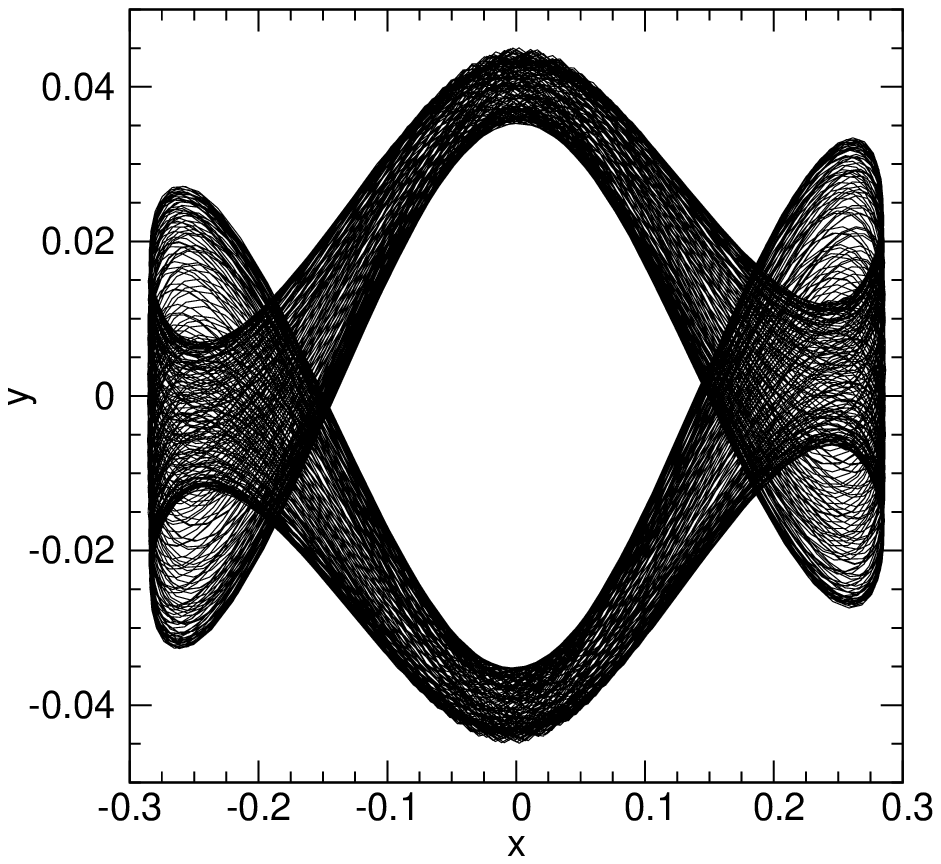}}
\caption{Two orbits that cross themselves: Orbits 1855 of model E2af (left)
and 142 of model E5af (right); notice that, for the latter, the scales of
the $y$ and $z$ axes are very different than that of the $x$ axis, because
the orbit is stronly elongated.} 
\label{cross}
\end{figure*}

Another interesting feature of our frequency maps is that they are
dominated by the diagonal $F_x/F_z \simeq F_y/F_z$, i.e., the $(1, -1, 0)$
resonance, that in non-rotating systems corresponds to the SATs. 
In fact, most of the orbits in those diagonals in Fig. \ref{fmap} are
also SATs: after excluding those with $F_y/F_z \simeq 1.0$ and the
horseshoes mentioned above, there
remain 857 in model E2af and 916 in model E5af, with 591 of the former
and 802 of the latter conserving the sign of the $z$ component
of the angular momentum, i.e., they are SATs as shown in the two
examples of Fig. \ref{sats}. Most of the rest, i.e., the orbits
that do not conserve the sign of the $z$ component of the angular
momentum, are also SATs that cross themselves simply because they
are in a rotating system, as shown on the left side of Fig.
\ref{cross}, but there are also some of the type shown, e.g.,
in figure 3.19 of \citet{BT08}, as shown on the right side of
our figure. A search for additional resonances,
with $|l|$, $|m|$ and $|n|$ not larger than 10, among the orbits on the
$F_{x}/F_{z} \simeq F_{y}/F_{z}$ line yielded only 22 in model E2af and 25 in
model E5af, most of them of high order.

\begin{figure*}
\resizebox{\hsize}{!}{\includegraphics{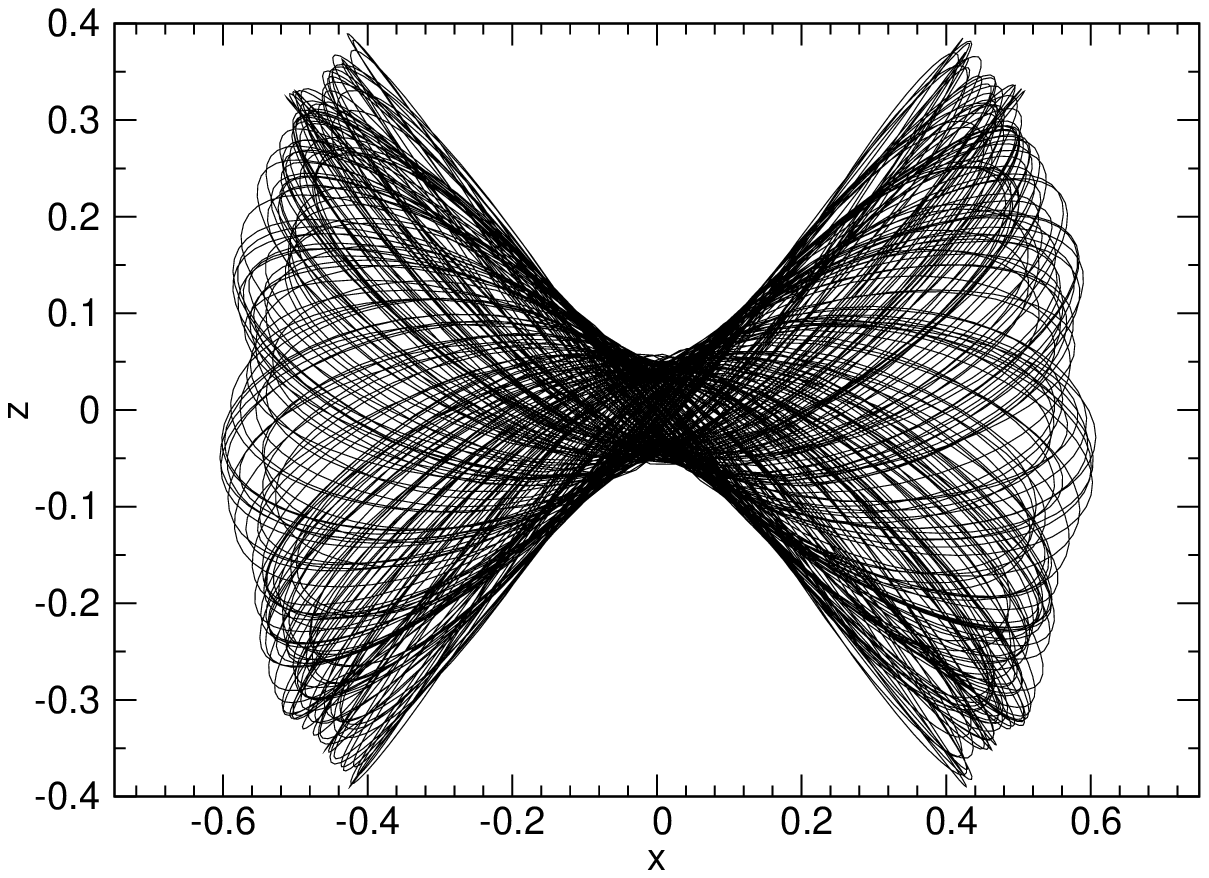}\hspace{1cm}
                      \includegraphics{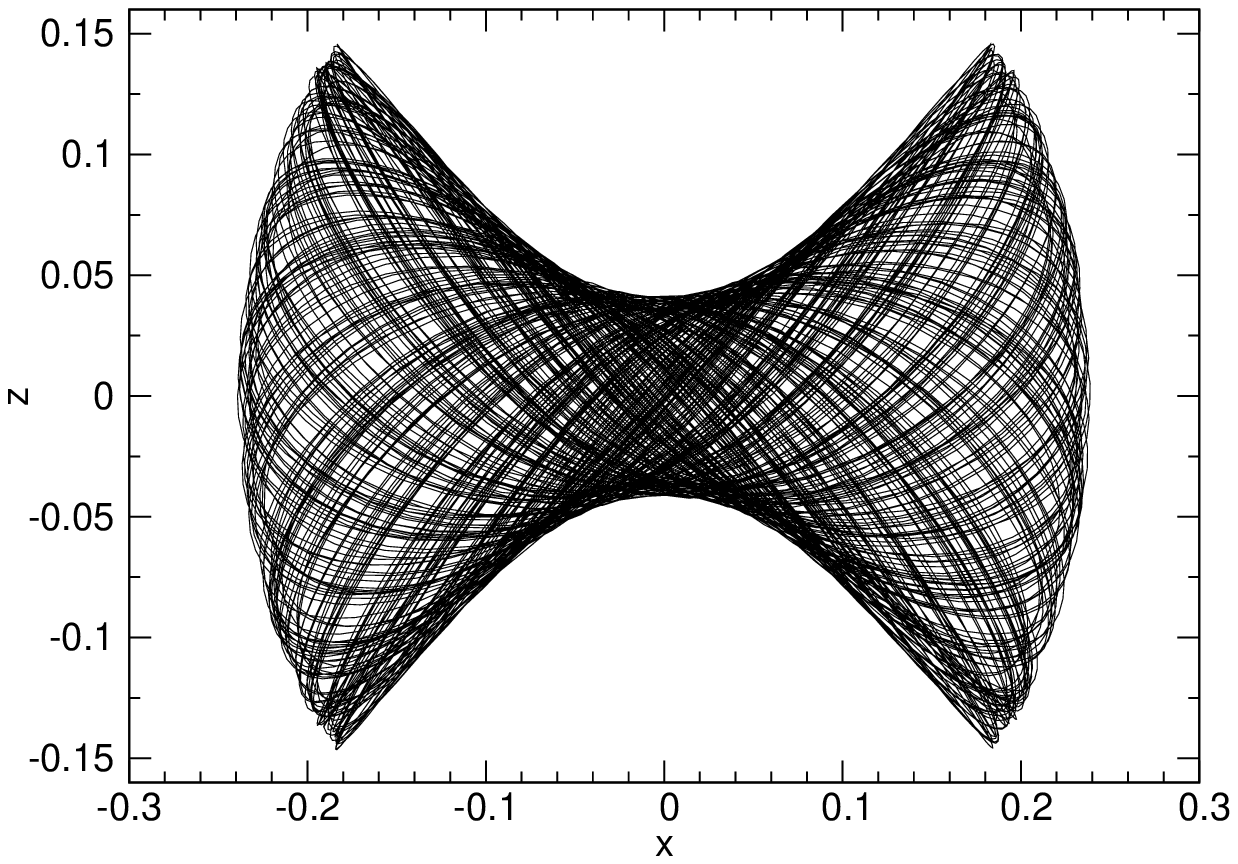}}
\resizebox{\hsize}{!}{\includegraphics{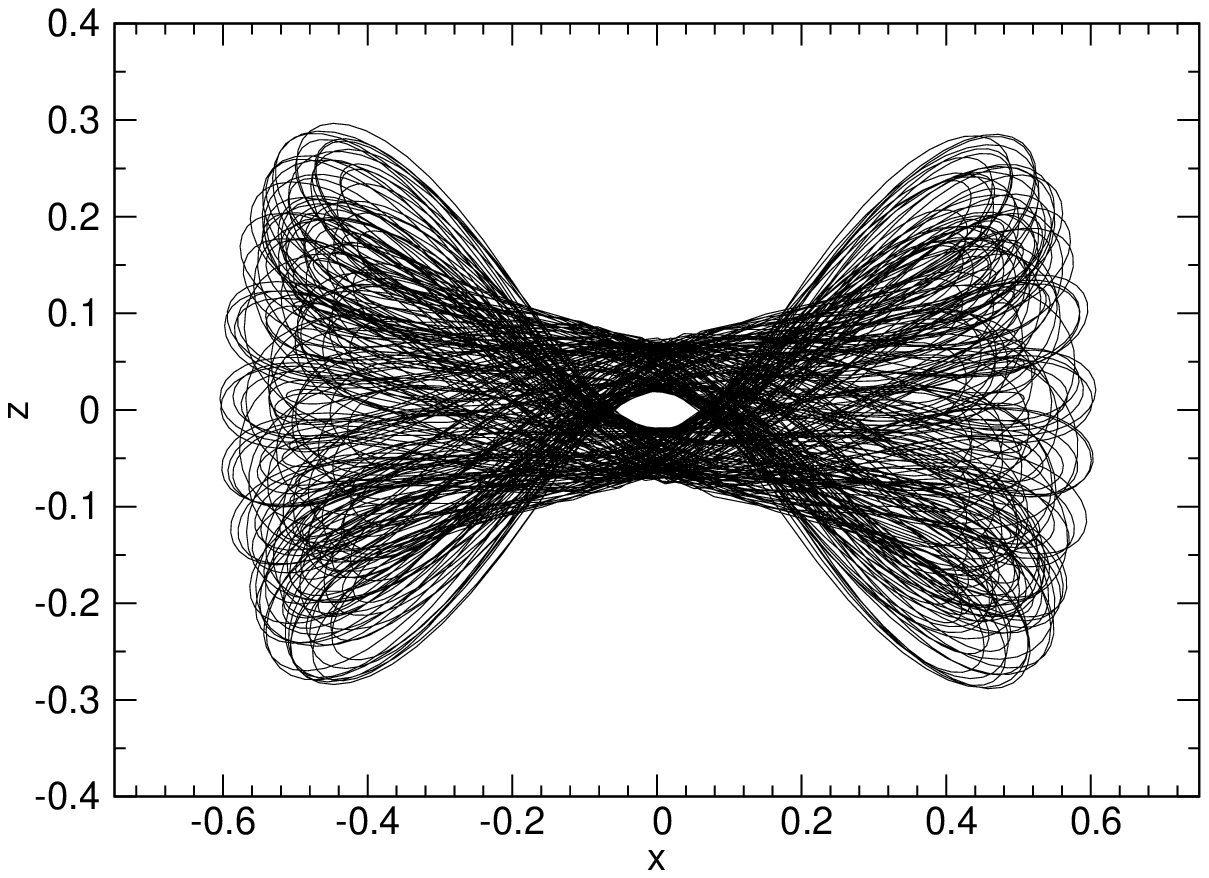}\hspace{1cm}
                      \includegraphics{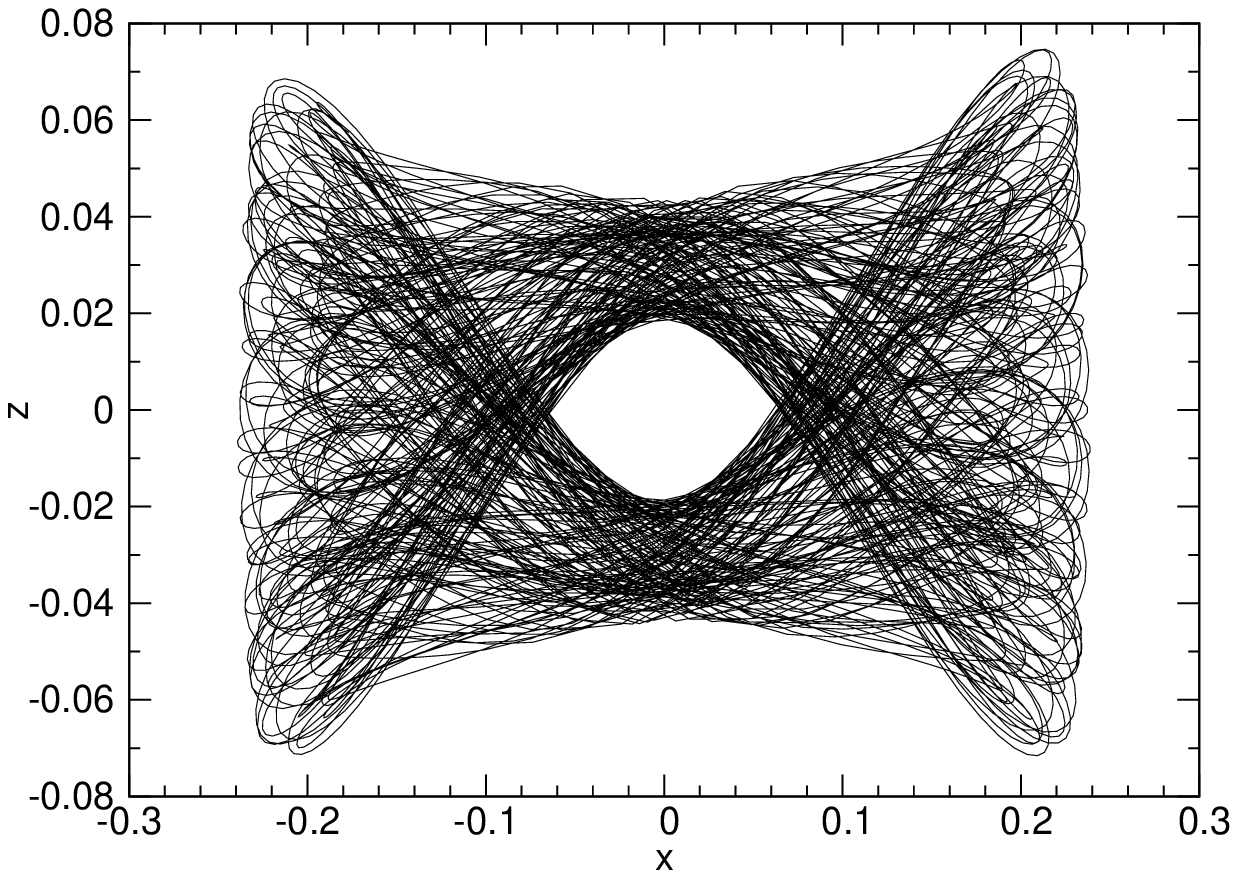}}
\caption{Two orbits lying outside the $(1,-1,0)$ resonance:
orbits 304 of model 
E2af (left) and 468 of model E5af (right).} 
\label{outres}
\end{figure*}

\begin{figure*}
\resizebox{\hsize}{!}{\includegraphics{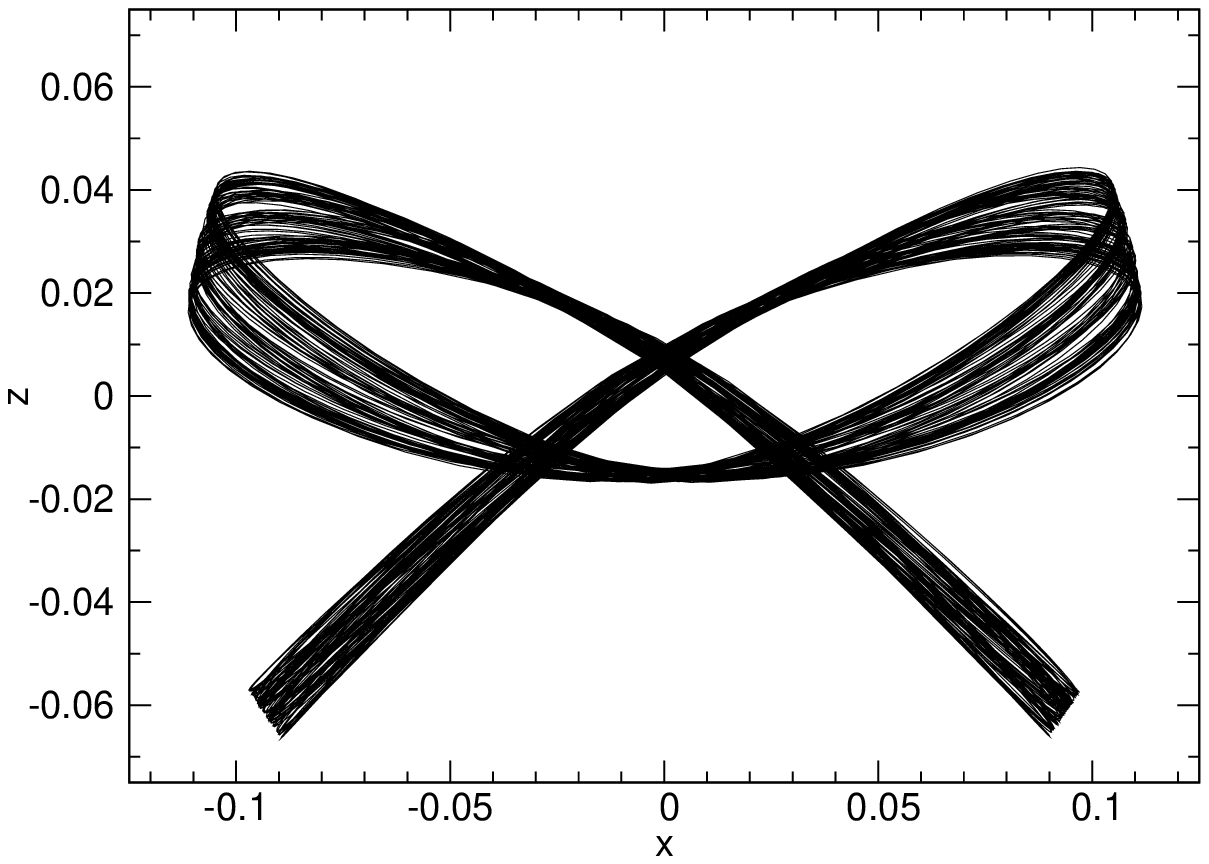}\hspace{1cm}
                      \includegraphics{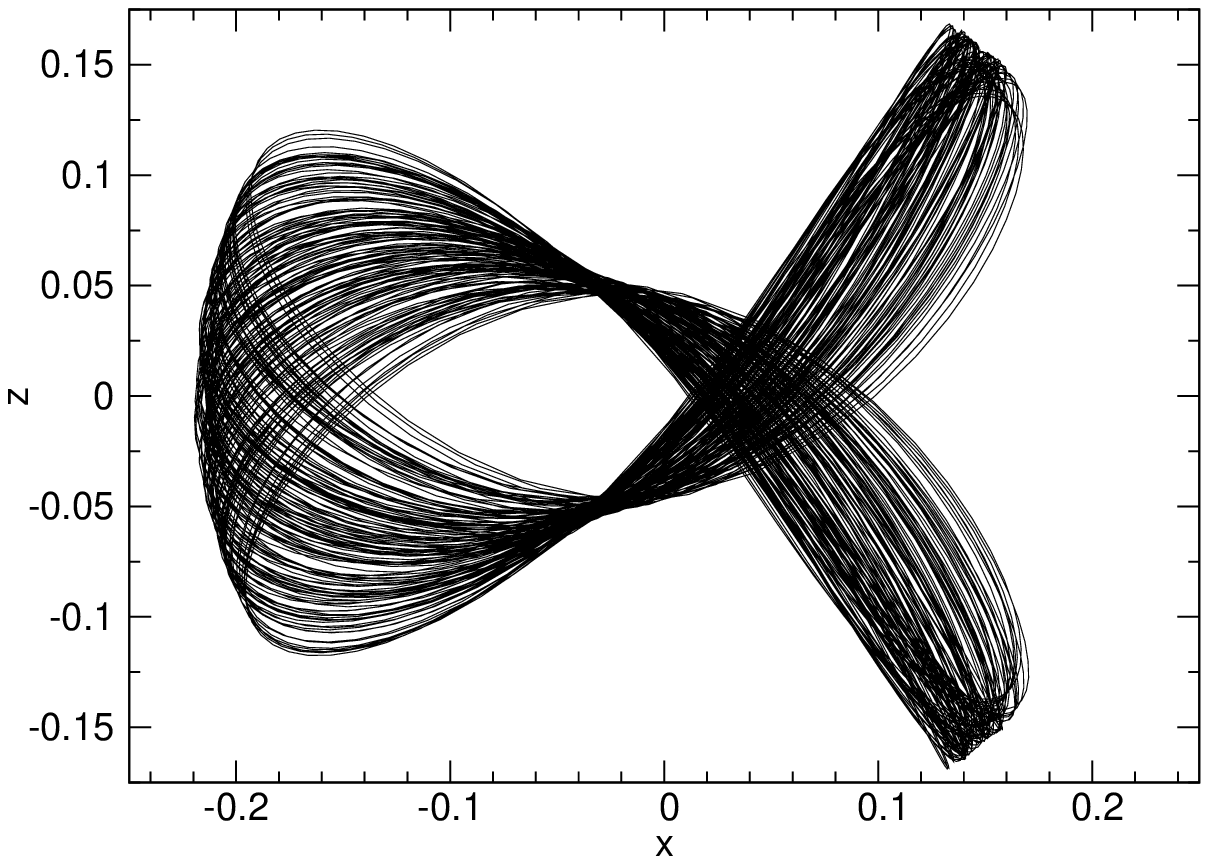}}
\resizebox{\hsize}{!}{\includegraphics{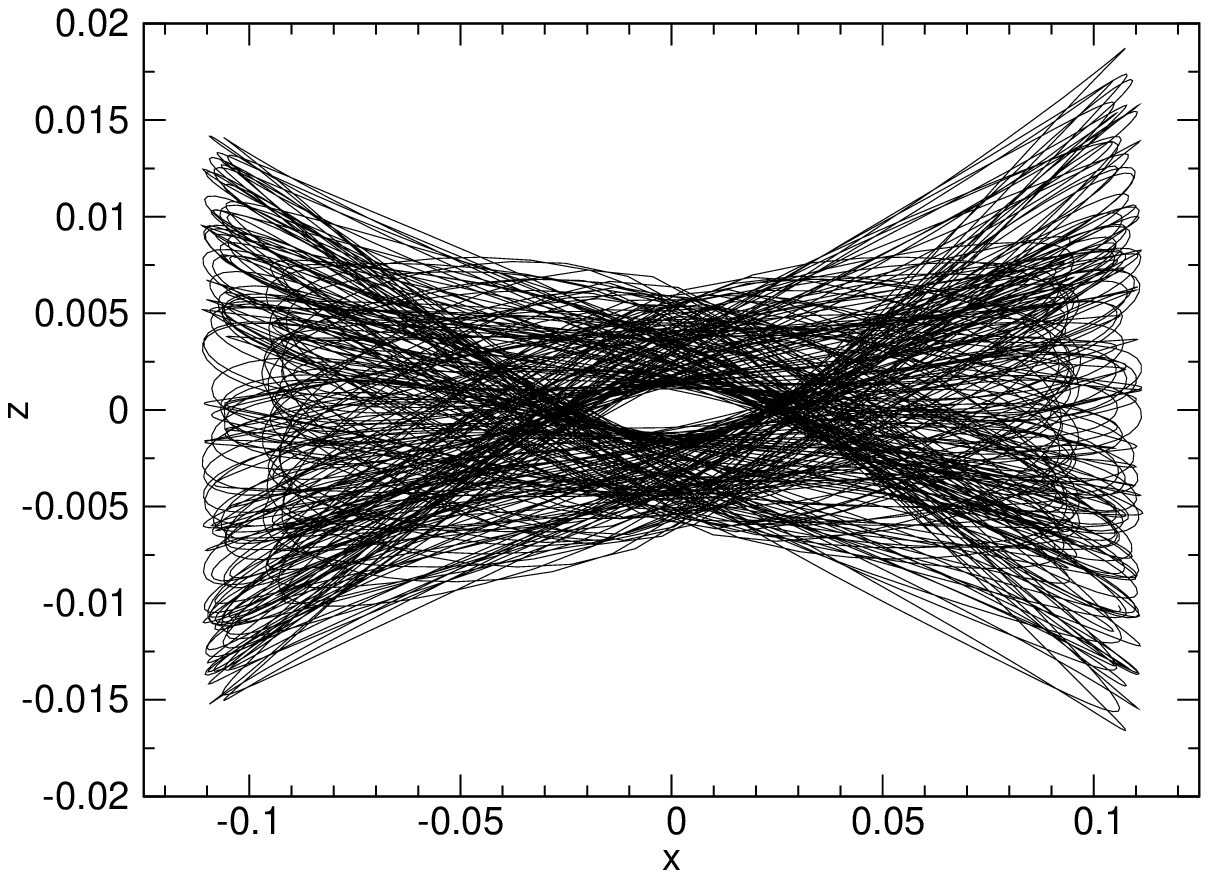}\hspace{1cm}
                      \includegraphics{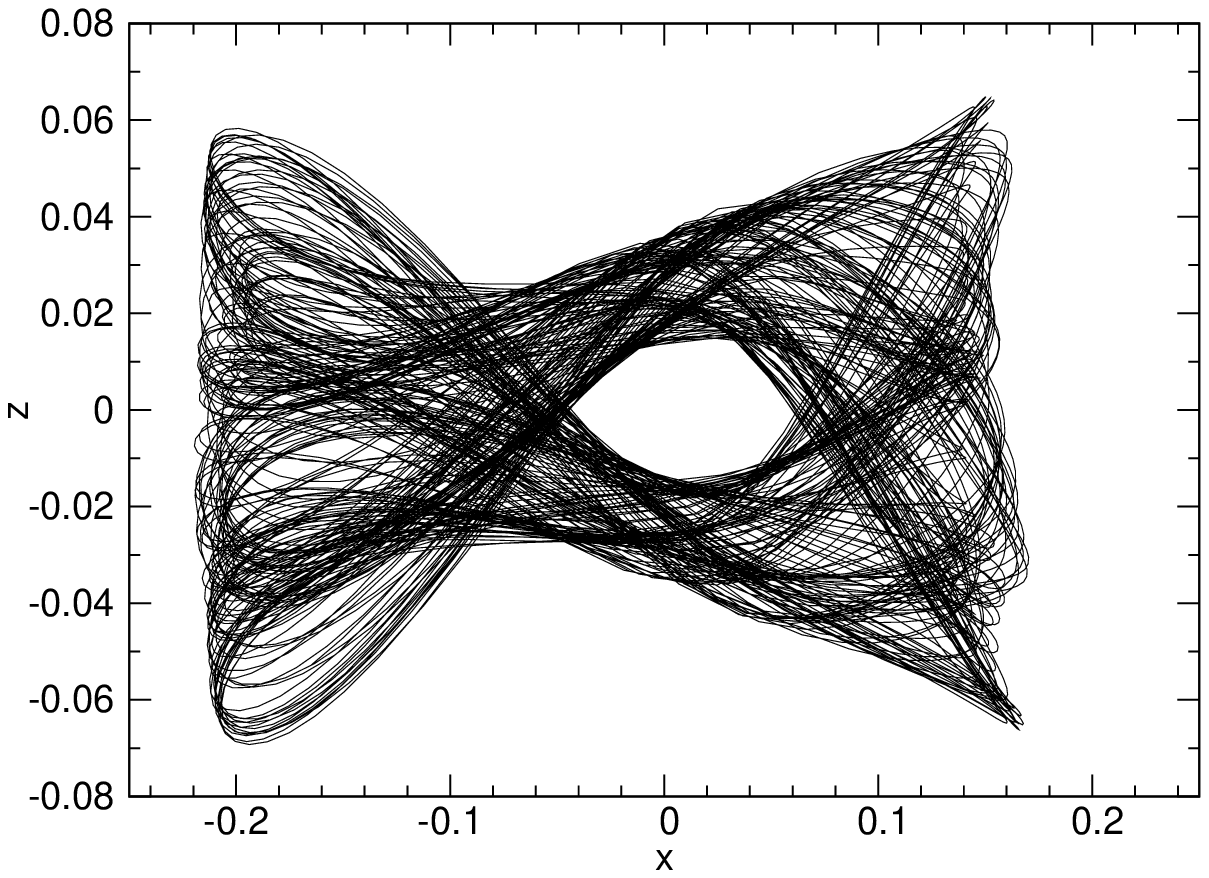}}
\caption{Two resonant orbits lying outside the $(1,-1,0)$
resonance: orbits 2692 of model 
E2af (left) and 2164 of model E5af (right); the former is a pretzel that obeys
the 
resonance $(4,0,-3)$, and the latter a fish that obeys the resonance $(3,0,-2)$.}
\label{pretzfish}
\end{figure*}

Except for the few cases with low $F_x/F_z$ values mentioned
above, the points falling outside the $F_x/F_z \simeq F_y/F_z$ lie in the region
occupied by the boxes and boxlets of non--rotating models \citep[see, e.g,
figure 1 of][]{MNZ13}, but most of them are more similar to orbits like that shown
on the right side of Fig. \ref{cross} than to boxes; two examples are shown in
Fig. \ref{outres}. A search for resonances, with $|l|$, $|m|$ and $|n|$ not larger
than 10, among these orbits found that 34 out of 63 of model E2f, and 26 out of 47
of model E5af obeyed at least one, but there does not seem to be any one
particularly frequent or relevant; two examples of these resonant orbits are shown
in Fig. \ref{pretzfish}.

\section{Conclusions}

We have shown here that it is possible, starting from a non-rotating self-consistent
$N$--body model of a triaxial and cuspy stellar system, to create a similar
rotating one by adding rotation to its particles and letting it to relax towards
a new equilibrium. The outcome is a system with differential rotation, with the angular
velocity depending both on the distance to the center and on the height above the
equatorial plane, but with a very uniform figure rotation. Besides,  these
models are highly stable over intervals of the order of a Hubble time.

\cite{ZM12} had shown that the density distribution of their
models mimics that of elliptical galaxies and our Figs. \ref{ge2a} and \ref{ge5a}
(upper left) show the same for the present models. The rotation of our models
can be compared with the results of \cite{Eea07} who define a global rotation
parameter $\lambda_R$ that, in their Appendix A, they related to our $\lambda$ by
\begin{equation}
\lambda\simeq \frac{\sqrt{2}}{3}\lambda_R.
\end{equation}
The corresponding values for our models E2af and E5af are, thus, $\lambda_R=0.263$
and $\lambda_R=0.378$, respectively, and one can find several examples of real
galaxies with similar values in the Table 1 of \cite{Eea07}. Besides, \cite{CCE04}
adjusted an axisymmetric model to SAURON \citep{B01} data on NGC 3377 and they computed the
velocity dispersions of their model at different points on the meridian plane.
Using their reported distance to NGC 3377 (9.9 Mpc) to transform their angular
distances into linear ones, and computing the anisotropy from the curves of their
figure 13, it turns out to be of the same order of magnitude than our own values.

To investigate figure rotation observations alone are not
enough and models have to be fitted to them. \cite{SEPB04} did this
for the old elliptical NGC 4365 and they found hints of figure rotation, roughly
estimating a period of about 5 Gyr, somewhat longer than for our models. It
should be noted, however, that their model rotates about the semimajor axis,
rather than the small axis as our models do. Interestingly, they found that
NGC 4365 is triaxial, with $T\simeq 0.45$, i.e., very similar to the triaxiality
of our model E2af ($T=0.447$).

About two thirds of the orbits in our models are
chaotic and bodies on regular, partially and fully chaotic orbits have different
spatial distributions. The elliptical galaxies that our chaotic models represent
would have rotation periods within $10^8 - 10^9$ years range, thus confirming
the suggestion of \cite{DVM11} that unstable orbits appear within that range.
Neverthelss, we cannot agree with their statement that stable systems
are unlikely to have rotation speeds that produce a high level
of stochasticity, as our own models prove otherwise. We had shown in our
previous investigations that non-, or very slowly, rotating triaxial stellar
models with very high fractions of chaotic orbits can be perfectly stable;
now the present work extends that conclusion to rotating models as well. 

Regular families are dominated by far by the SATs. The fractions of boxes and
boxlets are low as are those of horseshoes and orbits that cross themselves,
typical of rotating systems. The LATs
are replaced by tubes whose axes lie on the short--long axes
plane, but do not coincide with the major axis.

Finallly, it is interesting to compare our frequency maps with those of
\cite{DVM11}, who investigated sample orbits in a rotating triaxial \cite{D93}
potential, i.e, without considering self consistency. Thus, it is not surprising
that our Fig. \ref{fmap} shows significant empty spaces in regions where their
figures 2, 3, 7, and 10  are well populated. It is simply the result that not
all \emph{possible} orbits will be \emph{actually} present in a self-consistent
model, because self-consistency imposes a strong selection effect, a fact
beautifully proved by \cite{KV05}.

\section{Acknowledgements}

We are very grateful to L. Hernquist, D. Nesvorn\'y 
and D. Pfenniger for allowing us to use their codes, and to
R.E. Mart\'{\i}nez and to H.R Viturro for their technical assistance.
The comments of an anonymous referee were very useful
to improve the original version of this paper, and the assistance of M. Muzzio
to improve the English of that version is gratefully acknowledged.
This work was supported with grants from the
Consejo Nacional de Investigaciones Cient\'{\i}ficas y T\'ecnicas de la
Rep\'ublica Argentina, the Agencia Nacional de Promoci\'on Cient\'{\i}fica
y Tecnol\'ogica and the Universidad Nacional de La Plata.

\bsp

\label{lastpage}

\end{document}